\documentstyle[psfig]{mn}

\newif\ifAMStwofonts
\AMStwofontstrue

\ifoldfss
  \ifCUPmtlplainloaded \else
    \NewTextAlphabet{textbfit} {cmbxti10} {}
    \NewTextAlphabet{textbfss} {cmssbx10} {}
    \NewMathAlphabet{mathbfit} {cmbxti10} {} 
    \NewMathAlphabet{mathbfss} {cmssbx10} {} 
  \fi
  \ifAMStwofonts
    \ifCUPmtlplainloaded \else
      \NewSymbolFont{upmath} {eurm10}
      \NewSymbolFont{AMSa} {msam10}
      \NewMathSymbol{\upi}     {0}{upmath}{19}
      \NewMathSymbol{\umu}     {0}{upmath}{16}
      \NewMathSymbol{\upartial}{0}{upmath}{40}
      \NewMathSymbol{\leqslant}{3}{AMSa}{36}
      \NewMathSymbol{\geqslant}{3}{AMSa}{3E}

      \let\leq=\leqslant 
      \let\geq=\geqslant 
    \fi
  \fi
\fi 

\ifnfssone
  \newmathalphabet{\mathit}
  \addtoversion{normal}{\mathit}{cmr}{m}{it}
  \addtoversion{bold}{\mathit}{cmr}{bx}{it}
  \newmathalphabet{\mathbfit} 
  \addtoversion{normal}{\mathbfit}{cmr}{bx}{it}
  \addtoversion{bold}{\mathbfit}{cmr}{bx}{it}
  \newmathalphabet{\mathbfss} 
  \addtoversion{normal}{\mathbfss}{cmss}{bx}{n}
  \addtoversion{bold}{\mathbfss}{cmss}{bx}{n}
  \ifAMStwofonts
    \ifCUPmtlplainloaded \else
      \UseAMStwoboldmath
      \makeatletter
      \new@mathgroup\upmath@group
      \define@mathgroup\mv@normal\upmath@group{eur}{m}{n}
      \define@mathgroup\mv@bold\upmath@group{eur}{b}{n}
      \edef\UPM{\hexnumber\upmath@group}
      \new@mathgroup\amsa@group
      \define@mathgroup\mv@normal\amsa@group{msa}{m}{n}
      \define@mathgroup\mv@bold\amsa@group{msa}{m}{n}
      \edef\AMSa{\hexnumber\amsa@group}
      \makeatother
      \mathchardef\upi="0\UPM19
      \mathchardef\umu="0\UPM16
      \mathchardef\upartial="0\UPM40
      \mathchardef\leqslant="3\AMSa36
      \mathchardef\geqslant="3\AMSa3E

      \let\leq=\leqslant 
      \let\geq=\geqslant 
    \fi
  \fi
\fi 

\ifnfsstwo
  \DeclareMathAlphabet{\mathbfit}{OT1}{cmr}{bx}{it}
  \SetMathAlphabet\mathbfit{bold}{OT1}{cmr}{bx}{it}
  \DeclareMathAlphabet{\mathbfss}{OT1}{cmss}{bx}{n}
  \SetMathAlphabet\mathbfss{bold}{OT1}{cmss}{bx}{n}
  \ifAMStwofonts
    \ifCUPmtlplainloaded \else
      \DeclareSymbolFont{UPM}{U}{eur}{m}{n}
      \SetSymbolFont{UPM}{bold}{U}{eur}{b}{n}
      \DeclareSymbolFont{AMSa}{U}{msa}{m}{n}
      \DeclareMathSymbol{\upi}{0}{UPM}{"19}
      \DeclareMathSymbol{\umu}{0}{UPM}{"16}
      \DeclareMathSymbol{\upartial}{0}{UPM}{"40}
      \DeclareMathSymbol{\leqslant}{3}{AMSa}{"36}
      \DeclareMathSymbol{\geqslant}{3}{AMSa}{"3E}

      \let\leq=\leqslant 
      \let\geq=\geqslant 
    \fi
  \fi
\fi 

\ifCUPmtlplainloaded \else
  \ifAMStwofonts \else 
    \def\upi{\pi}
    \def\umu{\mu}
    \def\upartial{\partial}
  \fi
\fi

\title{Nova nucleosynthesis and Galactic evolution of the CNO isotopes}
\author[D. Romano and F. Matteucci]
       {Donatella Romano$^{1, 2}$ and Francesca Matteucci$^3$\\
        $^1$International School for Advanced Studies, SISSA/ISAS, 
	    Via Beirut 2-4, I-34014 Trieste, Italy\\
        $^2$INAF, Osservatorio Astronomico di Bologna, 
	    Via Ranzani 1, I-40127 Bologna, Italy; romano@bo.astro.it\\
	$^3$Dipartimento di Astronomia, Universit\`a di Trieste,
            Via G.B. Tiepolo 11, I-34131 Trieste, Italy; matteucci@ts.astro.it}
\date{Accepted .
      Received ;
      in original form }

\pagerange{\pageref{firstpage}--\pageref{lastpage}}
\pubyear{2003}

\begin{document}

\maketitle

\label{firstpage}

\begin{abstract}
In this paper we study the role played both by novae and single stars in 
enriching the interstellar medium of the Galaxy with CNO group nuclei, in the 
framework of a detailed successful model for the chemical evolution of both 
the Galactic halo and disc. First, we consider only the nucleosynthesis from 
single low-mass, intermediate-mass and massive stars. In particular, the 
nucleosynthesis prescriptions in the framework of the adopted model are such 
that: {\it i)} low- and intermediate-mass stars are responsible for the 
production of most of the Galactic $^{12}$C and $^{14}$N; {\it ii)} massive 
stars produce the bulk of the Galactic $^{16}$O; {\it iii)} $^{13}$C and 
$^{17}$O originate mostly in intermediate-mass stars, with only a minor 
contribution from low-mass and massive stars. In this context, we show that 
the behavior of the $^{12}$C/$^{13}$C, $^{14}$N/$^{15}$N and $^{16}$O/$^{17}$O 
isotopic ratios, as inferred from observations, can be explained only allowing 
for a substantial revision of the available stellar yields. On the other hand, 
the introduction of nova nucleosynthesis allows us to better explain the 
temporal evolution of the CNO isotopic ratios in the solar neighbourhood as 
well as their trends across the Galactic disc. Once all the nucleosynthesis 
sources of CNO elements are taken into account, we conclude that $^{13}$C, 
$^{15}$N and $^{17}$O are likely to have both a primary and a secondary 
origin, in contrast to previous beliefs. We show that, when adopting the 
most recent $^{17}$O yields from intermediate-mass stars published in the 
literature so far, we still get a too large solar abundance for this element, 
a problem already encountered in the past by other authors using different 
yield sets. Therefore, we conclude that in computing the $^{17}$O yields from 
intermediate-mass stars some considerable sink of $^{17}$O is probably 
neglected. The situation for $^{15}$N is less clear than that for $^{13}$C and 
$^{17}$O, mainly due to contradictory observational findings. However, a 
stellar factory restoring $^{15}$N on quite long time-scales seems to be 
needed in order to reproduce the observed positive gradient of 
$^{14}$N/$^{15}$N across the disc, and novae are, at present, the best 
candidates for this factory. Given the uncertainties still present in the 
computation of theoretical stellar yields, our results can be used to put 
constraints on stellar evolution and nucleosynthesis models.
\end{abstract}

\begin{keywords}
Galaxy: abundances -- Galaxy: evolution -- novae, cataclysmic variables -- 
nuclear reactions, nucleosynthesis, abundances
\end{keywords}

\section{Introduction}

In the last 30 years, many studies have been devoted to the evolution of 
the CNO isotopic ratios (e.g., Audouze, Lequeux \& Vigroux 1975; Vigroux, 
Audouze \& Lequeux 1976; Dearborn, Tinsley \& Schramm 1978; Tosi 1982; 
D'Antona \& Matteucci 1991; Matteucci \& D'Antona 1991; Prantzos, Aubert \& 
Audouze 1996; see also Tosi 2000 for a recent reappraisal of the problem). 
Isotopic ratios are generally not affected by physicochemical fractionation 
effects (but see, e.g., Sheffer, Lambert \& Federman 2002). Therefore, they 
reflect rather faithfully the relevant production processes which occurred in 
stars of different masses and lifetimes. However, despite the considerable 
progress in the theories of stellar evolution and nucleosynthesis, important 
questions related to the yields and the production sites of some of the CNO 
isotopes still remain open. In this context, chemical evolution models can be 
regarded as a powerful tool in order to discriminate among different sets of 
stellar yields and even different stellar factories. In fact, as discussed by 
Tosi (1988), model predictions on elemental and isotopic ratios depend 
primarily on the adopted stellar nucleosynthesis, stellar lifetimes and 
initial mass function (IMF), rather than on the other uncertain parameters of 
the galactic evolution, like star formation and infall rates.

Among the CNO group nuclei, $^{16}$O is the best understood. It is a primary 
element, i.e., it is always synthesized starting from H and He in the parent 
star. When only the most reliable oxygen measurements are considered, a good 
agreement is found between theoretical predictions on $^{16}$O evolution and 
observations (e.g., Chiappini, Romano \& Matteucci 2003), which confirms that 
$^{16}$O production is well understood both qualitatively and quantitatively. 
The bulk of $^{16}$O comes from massive stars ($m >$ 10 $M_\odot$).

The production of $^{12}$C is instead more uncertain: we know that it is 
synthesized as a primary element, but the exact amount restored into the 
interstellar medium (ISM) by stars of different masses is still uncertain. The 
`old' view that $^{12}$C is mainly produced by low- and intermediate-mass 
stars (LIMS) has been challenged by several papers (Prantzos, Vangioni-Flam \& 
Chauveau 1994; Carigi 2000; Henry, Edmunds \& K\"oppen 2000) suggesting that 
mass-loosing massive stars could be the main source of carbon. The conclusion 
of those authors rests on the adoption of the metallicity dependent yields for 
massive stars from Maeder (1992), which are computed in presence of mass loss. 
On the other hand, very recently it has been demonstrated that, in order to 
reproduce simultaneously the features of the C/O vs. O/H diagrams observed for 
galaxies of different morphological type, as well as the [C/Fe] vs. [Fe/H] and 
[C/O] vs. [Fe/H] diagrams for dwarf stars in the solar vicinity, the bulk of 
$^{12}$C should originate from LIMS (Chiappini et al. 2003). Indeed, the 
carbon yields due to mass loss have been revised downward (Meynet \& Maeder 
2002a).

The nature of $^{14}$N is still controversial: it should be a typical 
secondary element, but with a primary component as well. The secondary 
component is mainly produced through the CN cycle in stars of all masses, thus 
its amount depends on the $^{12}$C present in the star at the time of its 
birth. Secondary nitrogen can also be produced in the ON cycle by 
transformation of $^{16}$O, but at a much slower rate. The primary component 
is instead produced starting from fresh carbon generated by the parent star, 
provided there is some mixing mechanism which transports this newly 
synthesized carbon in a hydrogen-burning region, where the CNO cycle can 
convert it to nitrogen. In intermediate-mass stars (2.5\,--\,3.5 $\leq$ 
$m/M_\odot$ $\leq$ 6\,--\,8), primary $^{14}$N is produced when hot bottom 
burning (HBB) takes place at the base of the star convective envelope, during 
the asymptotic giant branch (AGB) phase (Renzini \& Voli 1981). The amount of 
primary nitrogen depends mostly on the adopted value of the mixing length 
parameter determining the depth of the convective stellar envelope, which is 
very uncertain (see a discussion in van den Hoek \& Groenewegen 1997). Primary 
$^{14}$N production might take place also in massive stars (e.g., Matteucci 
1986; Meynet \& Maeder 2002b). According to Meynet \& Maeder (2002b), the 
mechanism which allows the freshly made carbon to be turned into primary 
nitrogen is stellar axial rotation, and primary nitrogen production in 
massive stars is favored at low metallicity. In fact, in this case (rotating 
stellar models at low $Z$), the stars loose less angular momentum, thus they 
rotate faster. Simultaneously, they are more compact, so that differential 
rotation and shear mixing are stronger. Moreover, the H-burning shell has a 
much higher temperature and is thus closer to the core, which favours mixing 
between the two. However, the primary nitrogen yields obtained by Meynet \& 
Maeder from their low-metallicity massive star models do not contribute to a 
significant chemical enrichment at very low metallicities (see Chiappini et 
al. 2003).

As far as the minor isotopes are concerned, it has been claimed that $^{13}$C 
and possibly also $^{17}$O are mainly formed in the external regions of stars 
in the red giant branch (RGB), planetary nebula (PN) and supernova (SN) phase 
(Audouze et al. 1975; Dearborn et al. 1978). Marigo, Bressan \& Chiosi (1996) 
explored the formation of a {\it pocket} of primary $^{13}$C during the AGB 
phase of intermediate-mass stars.

The thermonuclear runaway (TNR) responsible for the nova outbursts has also 
been identified as a promising channel for the synthesis of $^{13}$C, 
$^{15}$N, and $^{17}$O (Starrfield et al. 1972; Starrfield, Sparks \& Truran 
1974), eventually being responsible for most of the Galactic abundances of 
$^{13}$C and $^{17}$O (Jos\'e \& Hernanz 1998; Woosley et al. 1997). Indeed, 
classical novae, binary systems consisting of a carbon-oxygen or 
oxygen-neon-magnesium white dwarf accreting hydrogen-rich matter from a 
main-sequence companion, sporadically inject nuclearly processed material into 
the ISM. The TNR, responsible for the explosion causing the ejection of almost 
the whole previously accreted envelope, leads to the synthesis of some rare 
nuclei during the nova outburst. The most recent nova hydrodynamical models 
confirm large overproduction factors with respect to the solar abundances for 
$^7$Li, $^{13}$C, $^{15}$N and $^{17}$O (Jos\'e \& Hernanz 1998). Therefore, 
although novae are processing only a minor fraction of the total matter in a 
galaxy, nevertheless they could produce important amounts of the species with 
the highest overproduction factors. In the past, it has been suggested that 
the bulk of $^{15}$N production should come from novae (Audouze et al. 1975; 
Dearborn et al. 1978; Matteucci \& D'Antona 1991); Type II SNe might also 
contribute, although to a lower extent (Audouze et al. 1977; Timmes, Woosley 
\& Weaver 1995). Recent detections of extragalactic $^{15}$N seem to support 
the idea that $^{15}$N production is due to rotationally induced mixing of 
protons into the helium-burning shells of massive stars (Chin et al. 1999 and 
references therein). A small contribution from Type Ia SNe could also be 
possible (Clayton et al. 1997; although see Nomoto, Thielemann \& Yokoi 1984).

The aim of this paper is to present some theoretical results concerning the 
evolution of the $^{12}$C/$^{13}$C, $^{14}$N/$^{15}$N, and $^{16}$O/$^{17}$O 
isotopic ratios in the solar neighbourhood as well as along the Galactic disc 
by taking advantage of a complete model for the chemical evolution of the 
Galaxy in which the nucleosynthesis from stars of different masses is included 
in a detailed way. In particular, we include in the model the nucleosynthesis 
from novae and try to establish whether they can actually be regarded as 
important contributors to the Galactic abundances of the CNO  isotopes 
$^{13}$C, $^{15}$N, and $^{17}$O. For each of these stable nuclei, we make an 
attempt in identifying the main contributors to the chemical enrichment by 
means of a comparison with the available observational data.

The paper is organized as follows. In Section~2 we report on the status of the 
observations. In Section~3 we describe the adopted chemical evolution model. 
In Section~4 we present model results. Finally, our conclusions are drawn in 
Section~5.

\section{Observations}

\subsection{Galactic $^{\bmath{12}}$C/$^{\bmath{13}}$C abundance ratio}

The Solar System $^{12}$C/$^{13}$C ratio is ($^{12}$C/$^{13}$C)$_\odot$ = 89 
$\pm$ 2 (Cameron 1982), higher than the estimates for the local ISM: 
($^{12}$C/$^{13}$C)$_{\mathrm{ISM}}$ = 47.3$^{+\,5.5}_{-\,4.4}$ (from CN; 
Crane \& Hegyi 1988); 77 $\pm$ 3 (from CH$^+$; Stahl et al. 1989); 62 $\pm$ 4 
(from CO; Langer \& Penzias 1993); 77 $\pm$ 7 (average from H$_2$CO and CO 
data for sources near the Sun; Wilson \& Rood 1994); 58 $\pm$ 12 (from atomic 
C; Keene et al. 1998); 58 $\pm$ 6 (from atomic C$^+$; Boreiko \& Betz 1996); 
69 $\pm$ 15 (from the solid state CO$_2$; Boogert et al. 2000)\footnote{ A 
value of ($^{12}$C/$^{13}$C)$_{\mathrm{ISM}}$ between 60 and 80 is the 
preferred one.}. This clearly indicates a decrease in the last 4.5 Gyr, which 
is exactly what one expects for a primary to secondary elemental ratio 
(according to Talbot \& Arnett 1974, a chemical species is defined as primary 
or secondary according to whether the yields, i.e., the mass fractions of a 
star ejected in the form of the newly produced element, are insensitive, or 
sensitive, to the original metallicity of the star; as a consequence, the 
abundance of a secondary element increases more steeply in time than that of 
its seed nuclei).

The Galactic $^{12}$C/$^{13}$C gradient gives a consistent picture: 
$\Delta$($^{12}$C/$^{13}$C)/$\Delta$R$_G$ = (4.5 $\pm$ 2.2) kpc$^{-1}$ 
(weighted fit to the $^{12}$CO$_2$/$^{13}$CO$_2$ ratio as a function of 
Galactocentric radius -- Boogert et al. 2000). The gradient found from the 
solid state agrees well with gas phase studies (Tosi 1982; Wilson \& Rood 
1994). However, solid state determinations are more reliable than gas phase 
studies, because the column densities can be determined without uncertain 
radiative transfer effects. A limitation of the Boogert et al. analysis is the 
paucity of solid CO$_2$ observations at low Galactic radii (3\,--\,6 kpc). 
Moreover, the considerable scatter in the $^{12}$C/$^{13}$C ratio at radii 
larger than 6 kpc prevents from identifying a clear trend in the outer disc. 
However, the decrease of the ratio towards the Galactic centre is apparent and 
can be taken as a clear signature of an important secondary component to 
$^{13}$C nucleosynthesis. This decrease is confirmed also by $^{12}$C/$^{13}$C 
ratios derived from $^{12}$CN, $^{13}$CN observations, which however tend to 
be lower than those derived from other molecules (Savage, Apponi \& Ziurys 
2001).

\subsection{Galactic $^{\bmath{14}}$N/$^{\bmath{15}}$N abundance ratio}

The $^{14}$N/$^{15}$N ratio in our Galaxy has been studied by Dahmen, Wilson 
\& Matteucci (1995), who analysed high quality data of HCN double isotopomer 
ratios for 11 warm, dense molecular clouds in the Galactic disc. After fitting 
$^{12}$C/$^{13}$C ratios from published data as a function of Galactocentric 
distance, the $^{14}$N/$^{15}$N ratios were obtained. A linear regression 
using these ratios and previously published data allowed the determination of 
the Galactic gradient: $\Delta$($^{14}$N/$^{15}$N)/$\Delta$R$_G$ = (19.7 $\pm$ 
8.9) kpc$^{-1}$, shallower than that found by Tosi (1982).

The Solar System $^{14}$N/$^{15}$N ratio is ($^{14}$N/$^{15}$N)$_\odot$ = 270 
(Anders \& Grevesse 1989). For the local ISM, values as high as 
($^{14}$N/$^{15}$N)$_{\mathrm{ISM}}$ = 450 $\pm$ 22 (Wilson \& Rood 1994) or 
as low as ($^{14}$N/$^{15}$N)$_{\mathrm{ISM}}$ = 237$^{+\,27}_{-\,21}$ (Lucas 
\& Liszt 1998) have been reported. The former is in contradiction with the 
existence of a positive gradient across the Galactic disc since, from a 
theoretical point of view, if $^{14}$N/$^{15}$N increases in time and hence 
with metallicity, it should also increase with decreasing Galactocentric 
distance, owing to the higher metallicity content attained in the innermost 
Galactic regions.

\subsection{Galactic $^{\bmath{16}}$O/$^{\bmath{17}}$O abundance ratio}

The $^{16}$O/$^{17}$O abundance ratio has decreased from its solar value of 
($^{16}$O/$^{17}$O)$_\odot$ = 2465 (Anders \& Grevesse 1989) to its present 
local value of ($^{16}$O/$^{17}$O)$_{\mathrm{ISM}}$ = 1900 $\pm$ 200, an 
unmistakable sign of secondary production for $^{17}$O.

The secondary origin of $^{17}$O is confirmed by the behavior of the 
$^{16}$O/$^{17}$O gradient along the Galactic disc, which decreases with 
decreasing distance from the Galactic centre ($^{16}$O/$^{17}$O ratios along 
the Galactic disc are obtained by multiplying the corresponding 
$^{16}$O/$^{18}$O ratios by the ISM value of $^{18}$O/$^{17}$O = 3.5 -- see 
Prantzos et al. 1996 and references therein).

\section{The chemical evolution model}

\subsection{Basic aspects}

The {\it two-infall model} for the chemical evolution of the Galaxy 
(Chiappini, Matteucci \& Gratton 1997; Chiappini, Matteucci \& Romano 2001) is 
adopted, with the aim of reassessing the problem of whether novae 
significantly contribute to the Galactic chemical evolution of the minor CNO 
isotopes.

In the framework of the adopted model, the Galaxy forms out of two main infall 
episodes. During the first one, the bulge, inner halo and thick disc are 
built up, on a very short time-scale ($\sim$ 1 Gyr). During the second one, 
the thin disc is formed, on a much longer time-scale (being $\sim$ 7 Gyr at 
the solar ring and increasing with increasing Galactocentric distance) out of 
matter of primordial chemical composition plus traces of halo gas.

For a complete description of the model basic assumptions and equations we 
address the interested reader to Chiappini et al. (1997, 2001, 2003). Here we 
only briefly outline how novae have been included in the model (see also 
D'Antona \& Matteucci 1991 and Romano et al. 1999). The rate of formation of 
nova systems at a given time $t$ is computed as a fraction $\alpha$ of the 
rate of formation of white dwarfs (WDs) at a previous time $t - \Delta t$:

\begin{equation}
R_{\mathrm{novae}}(t) = \alpha \int_{0.8}^8 \psi(t - \tau_m - \Delta t) 
                        \varphi(m) {\mathrm d}m,
\end{equation}
where $\tau_m$ is the lifetime of the star of mass $m$, $\psi(t)$ is the star 
formation rate and $\varphi(m)$ is the IMF. The Scalo (1986) IMF is assumed. 
All the stars between $m$ = 0.8 $M_\odot$ and $m$ = 8 $M_\odot$ end up as WDs. 
$\Delta t$ is a suitable delay-time which guarantees the cooling of the WD at 
a level that ensures a strong enough nova outburst. Here we assume $\Delta t$ 
= 2 Gyr, which is a typical value for novae (see also Romano et al. 1999 and 
D'Antona \& Mazzitelli 1982). The parameter $\alpha$ represents the fraction 
of WDs which belong to nova systems and its value (constant in time) is fixed 
by the request of reproducing the current rate of nova outbursts in the 
Galaxy. Unfortunately, this is a free parameter. Here we set $\alpha 
\sim$~0.01. The rate of nova outbursts is computed from the rate of formation 
of nova systems by assuming that each nova suffers roughly 10$^4$ outbursts 
during its life (Bath \& Shaviv 1978). A value of $\alpha$ between $\sim$~0.01 
and $\sim$~0.02 leads to $R_{\mathrm{outbursts}}(t_{\mathrm{Gal}})$ $\sim$ 
20\,--\,30 yr$^{-1}$, which has to be compared with the value inferred from 
scalings from extragalactic nova surveys, 
$R_{\mathrm{outbursts}}^{\mathrm{obs}}(t_{\mathrm{Gal}})$ = 15\,--\,50 
yr$^{-1}$ (Della Valle \& Livio 1994; see also Shafter 1997).

Since the evolution of the CNO isotopic ratios (as well as that of the other 
elemental ratios) depends mainly on the nucleosynthesis prescriptions, rather 
than on the parameters of the galactic evolution like star formation and 
infall rates, chemical evolution models can put constraints mainly on the 
nucleosynthesis and time-scales for element production. The validity of this 
statement rests on the fact that any variation in the star formation and/or 
infall laws affects the main and the minor isotopes {\it at the same extent}. 
In the framework of a model where nova nucleosynthesis is taken into account, 
the situation is more complicated. In fact, in this case the evolution of the 
CNO isotopic ratios depends also on the parameters which enter the computation 
of the theoretical nova outburst rate. Since novae produce large amounts of 
the {\it minor isotopes}, whereas their contribution to the main ones is 
almost negligible, each variation in the parameters regulating the nova rate 
does not cancel out in the CNO isotopic ratios. One should always keep in mind 
this when discussing predictions from models in which nova nucleosynthesis is 
included. We will come back on this issue in Section~4.3.

\subsection{Nucleosynthesis prescriptions}

The nucleosynthesis prescriptions adopted here are from {\it i)} van den Hoek 
\& Groenewegen (1997) and Ventura, D'Antona \& Mazzitelli (2002) for LIMS; 
{\it ii)} either Woosley \& Weaver (1995) or Nomoto et al. (1997) for Type II 
SNe; {\it iii)} Thielemann et al. (1993) for Type Ia SNe; {\it iv)} Jos\'e \& 
Hernanz (1998) for novae.

It is worth emphasizing that metallicity dependent yields are available from a 
limited number of studies, especially in the range of massive stars ($m >$ 10 
$M_\odot$). Moreover, nucleosynthesis studies are usually restricted to 
specific mass ranges (e.g., $m \sim$ 0.9\,--\,8.0 $M_\odot$ -- van den Hoek 
\& Groenewegen 1997; $m \sim$ 2.5\,--\,6.0 $M_\odot$ -- Ventura et al. 2002; 
$m \sim$ 11\,--\,40 $M_\odot$ -- Woosley \& Weaver 1995; $m \sim$ 13\,--\,70 
$M_\odot$ -- Nomoto et al. 1997) and/or do not deal with some specific 
chemical species, so that it is neither possible to homogeneously cover the 
mass spectrum over which stars distribute ($m \sim$ 0.1\,--\,100 $M_\odot$) 
nor to treat all the relevant chemical species by adopting a single stellar 
nucleosynthesis study.

For instance, van den Hoek \& Groenewegen (1997) give metallicity dependent 
yields of $^4$He, $^{12}$C, $^{13}$C, $^{14}$N and $^{16}$O for stars in the 
mass range $m \sim$ 0.9\,--\,8.0 $M_\odot$, but do not provide the yields of 
$^{17}$O. Therefore, one must complete their grid of stellar yields by means 
of $^{17}$O yields coming from some other study. We adopt the $^{17}$O yields 
recently computed by Ventura et al. (2002) for stars in the mass range $m 
\sim$ 2.5\,--\,6.0 $M_\odot$, for different initial chemical compositions. 
Since their yields do not extend to metallicities higher than $Z$ = 0.01, we 
must extrapolate them to supersolar metallicities. This makes the modelisation 
of $^{17}$O evolution in the inner disk inaccurate. Stars with $m <$ 2.0 
$M_\odot$ are not $^{17}$O producers, since they do not go through the NO 
cycle during which this element is synthesized. Therefore, we set to zero the 
$^{17}$O yields from stars in the mass range $m~\sim$~1.0\,--\,2.0 $M_\odot$. 
We also set to zero the $^{15}$N production from LIMS. In fact, the net yield 
of this element from LIMS is negative, due to the effect of the first and 
second dredge-up for stars in the mass range 0.8\,--\,3.5 $M_\odot$ and to HBB 
for higher mass stars (Marigo 2001).

Among the input parameters of nova hydrodynamical models with a deep influence 
on the nova nucleosynthesis there is the chemical composition of the H-rich 
envelope accreted by the WD from the main-sequence companion which fills its 
Roche lobe. The problem of the chemical composition of nova envelopes is 
complex and far from being understood. The matter transferred from the 
companion is assumed to be solarlike and is mixed in a given fraction with the 
outermost shells of the underlying core (e.g., Politano et al. 1995; Jos\'e \& 
Hernanz 1998). This is done in order to get the enhanced CNO or ONeMg 
abundances required both to power the explosion and to account for the 
spectroscopic abundance determination. It is not clear how the composition of 
the nova ejecta changes during the evolution of the Galaxy, as a function, 
e.g., of the chemical composition of the matter accreted from the WD 
companion. Dearborn et al. (1978) assumed that the amount of $^{15}$N ejected 
per nova outburst scales with the initial $^{12}$C\,+\,$^{13}$C\,+\,$^{14}$N 
abundance of the star. This means that $^{15}$N from novae is an element of 
purely secondary origin. On the contrary, $^{13}$C and $^{15}$N from novae 
were assumed to be of purely primary origin by D'Antona \& Matteucci (1991) 
and Matteucci \& D'Antona (1991). However, as noticed by these authors, a 
primary element mainly produced in long-lived progenitors still behaves like a 
secondary one from the point of view of galactic chemical evolution, owing to 
the delay with which it is restored to the ISM (see also Matteucci \& Fran\c 
cois 1989). In Section~4.3 we will show results obtained by assuming two 
limiting cases for $^{13}$C, $^{15}$N and $^{17}$O production during nova 
outbursts: in the first case, the yields of $^{13}$C, $^{15}$N and $^{17}$O 
from novae are assumed to be always the same, at each time of Galaxy's 
evolution; in the second case, they are scaled with the abundance of the seed 
nuclei in the star. 
\begin{table*}
\begin{minipage}{118mm}
\caption{Nucleosynthesis prescriptions.}
\begin{tabular}{c c c c c c}
\hline
Model & element & LIMS & massive stars$^{\mathrm{a}}$ & novae: 
ejecta$^{\mathrm{b}}$ & novae: s/p\\
\hline
 1 & $^{12}$C & vdHG97 ($\eta_{\mathrm{const}}$) & N97 & no & -- \\
   & $^{13}$C & vdHG97 ($\eta_{\mathrm{const}}$) & N97 & no & -- \\
   & $^{16}$O & vdHG97 ($\eta_{\mathrm{const}}$) & N97 & no & -- \\
   & $^{17}$O & VDM02/6 & N97 & no & -- \\
 & & & & & \\
 2 & $^{12}$C & vdHG97 ($\eta_{\mathrm{const}}$) & N97 & 
                       6.7 $\times$ 10$^{-3}$ $M_\odot$ & p \\
   & $^{13}$C & no & no & 1.2 $\times$ 10$^{-2}$ $M_\odot$ & p \\
   & $^{14}$N & vdHG97 ($\eta_{\mathrm{const}}$) & N97 & 
		1.7 $\times$ 10$^{-2}$ $M_\odot$ & p \\
   & $^{15}$N & no & no & 1.4 $\times$ 10$^{-3}$ $M_\odot$ & p \\
   & $^{16}$O & vdHG97 ($\eta_{\mathrm{const}}$) & N97 & 
		2.5 $\times$ 10$^{-1}$ $M_\odot$ & p \\
   & $^{17}$O & no & no & 1.4 $\times$ 10$^{-3}$ $M_\odot$ & p \\
 & & & & & \\
 2{\it s} & $^{12}$C & vdHG97 ($\eta_{\mathrm{const}}$) & N97 & 
                6.7 $\times$ 10$^{-3}$ $M_\odot$ & p \\
   & $^{13}$C & no & no & 1.5 $\times$ 10$^{-2}$ $M_\odot$ & s \\
   & $^{14}$N & vdHG97 ($\eta_{\mathrm{const}}$) & N97 & 
		1.7 $\times$ 10$^{-2}$ $M_\odot$ & p \\
   & $^{15}$N & no & no & 3.8 $\times$ 10$^{-3}$ $M_\odot$ & s \\
   & $^{16}$O & vdHG97 ($\eta_{\mathrm{const}}$) & N97 & 
		2.5 $\times$ 10$^{-1}$ $M_\odot$ & p \\
   & $^{17}$O & no & no & 1.8 $\times$ 10$^{-3}$ $M_\odot$ & s \\
 & & & & & \\
 3 & $^{12}$C & vdHG97 ($\eta_{\mathrm{const}}$) & N97 & 
		6.7 $\times$ 10$^{-3}$ $M_\odot$ & p \\
   & $^{13}$C & vdHG97 ($\eta_{\mathrm{const}}$) & N97 & 
		5.3 $\times$ 10$^{-3}$ $M_\odot$ & p \\
   & $^{14}$N & vdHG97 ($\eta_{\mathrm{const}}$) & N97 & 
		1.7 $\times$ 10$^{-2}$ $M_\odot$ & p \\
   & $^{15}$N & no & no & 1.4 $\times$ 10$^{-3}$ $M_\odot$ & p \\
   & $^{16}$O & vdHG97 ($\eta_{\mathrm{const}}$) & N97 & 
		2.5 $\times$ 10$^{-1}$ $M_\odot$ & p \\
   & $^{17}$O & VDM02/10.5 & N97 & 6.4 $\times$ 10$^{-4}$ $M_\odot$ & p \\
 & & & & & \\
 3{\it s} & $^{12}$C & vdHG97 ($\eta_{\mathrm{const}}$) & N97 & 
		6.7 $\times$ 10$^{-3}$ $M_\odot$ & p \\
   & $^{13}$C & vdHG97 ($\eta_{\mathrm{const}}$) & N97 & 
		1.5 $\times$ 10$^{-2}$ $M_\odot$ & s \\
   & $^{14}$N & vdHG97 ($\eta_{\mathrm{const}}$) & N97 & 
		1.7 $\times$ 10$^{-2}$ $M_\odot$ & p \\
   & $^{15}$N & no & no & 3.8 $\times$ 10$^{-3}$ $M_\odot$ & s \\
   & $^{16}$O & vdHG97 ($\eta_{\mathrm{const}}$) & N97 & 
		2.5 $\times$ 10$^{-1}$ $M_\odot$ & p \\
   & $^{17}$O & VDM02/10.5 & N97 & 8.8 $\times$ 10$^{-4}$ $M_\odot$ & s \\
 & & & & & \\
 3{\it n} & $^{12}$C & vdHG97 ($\eta_{\mathrm{var}}$) & WW95 & 
		       6.7 $\times$ 10$^{-3}$ $M_\odot$ & p \\
   & $^{13}$C & vdHG97 ($\eta_{\mathrm{var}}$) & WW95 & 
		1.5 $\times$ 10$^{-2}$ $M_\odot$ & s \\
   & $^{14}$N & vdHG97 ($\eta_{\mathrm{var}}$) & WW95 & 
		1.7 $\times$ 10$^{-2}$ $M_\odot$ & p \\
   & $^{15}$N & no & no & 3.8 $\times$ 10$^{-3}$ $M_\odot$ & s \\
   & $^{16}$O & vdHG97 ($\eta_{\mathrm{var}}$) & WW95 & 
		2.5 $\times$ 10$^{-1}$ $M_\odot$ & p \\
   & $^{17}$O & VDM02/10.5 & WW95 & 8.8 $\times$ 10$^{-4}$ $M_\odot$ & s \\
 & & & & & \\
 3{\it m} & $^{12}$C & vdHG97 ($\eta_{\mathrm{var}}$) & WW95 & 
		       6.7 $\times$ 10$^{-3}$ $M_\odot$ & p \\
   & $^{13}$C & vdHG97 ($\eta_{\mathrm{var}}$) & WW95 & 
		1.5 $\times$ 10$^{-2}$ $M_\odot$ & s \\
   & $^{14}$N & vdHG97 ($\eta_{\mathrm{var}}$) & WW95 & 
		1.7 $\times$ 10$^{-2}$ $M_\odot$ & p \\
   & $^{15}$N & no & no & 3.8 $\times$ 10$^{-3}$ $M_\odot$ & s \\
   & $^{16}$O & vdHG97 ($\eta_{\mathrm{var}}$) & WW95 & 
		2.5 $\times$ 10$^{-1}$ $M_\odot$ & p \\
   & $^{17}$O & VDM02/15 & WW95 & 8.8 $\times$ 10$^{-4}$ $M_\odot$ & s \\
\hline
\end{tabular}

$^{\mathrm{a}}$The yields of $^{12}$C have been multiplied by a factor of 
	       3 in the range 40\,--\,100 $M_\odot$ (see arguments in 
	       Chiappini et al. 2003).\\
$^{\mathrm{b}}$The mean masses ejected in form of $^{12}$C, $^{13}$C, 
	       $^{14}$N, $^{15}$N, $^{16}$O and $^{17}$O by each nova 
	       during its overall lifetime are: 6.7 $\times$ 10$^{-3}$ 
	       $M_\odot$, 1.5 $\times$ 10$^{-2}$ $M_\odot$, 1.7 $\times$ 
	       10$^{-2}$ $M_\odot$, 5.2 $\times$ 10$^{-3}$ $M_\odot$, 
	       2.5 $\times$ 10$^{-1}$ $M_\odot$ and 3.2 $\times$ 10$^{-3}$ 
	       $M_\odot$, respectively. These quantities are obtained by 
	       averaging the yields over the 14 evolutionary sequences 
	       given by Jos\'e \& Hernanz (1998) and by assuming that the 
	       mean mass ejected in a single burst is $\sim$ 2 $\times$ 
	       10$^{-5}$ $M_\odot$. They refer to $Z = Z_\odot$, and are 
	       computed by taking into account that 30\% of nova systems 
	       contain ONe WDs, while the remaining contain CO WDs. The 
	       quantities listed in column 5 are those actually needed in 
	       order to reproduce the solar abundances of $^{13}$C, 
	       $^{15}$N and $^{17}$O as given by Anders \& Grevesse (1989).
	       It is seen that in many cases it is necessary to lower the 
	       quantities given by the average described above.
\end{minipage}
\end{table*}
\begin{figure*}
\centerline{\psfig{figure=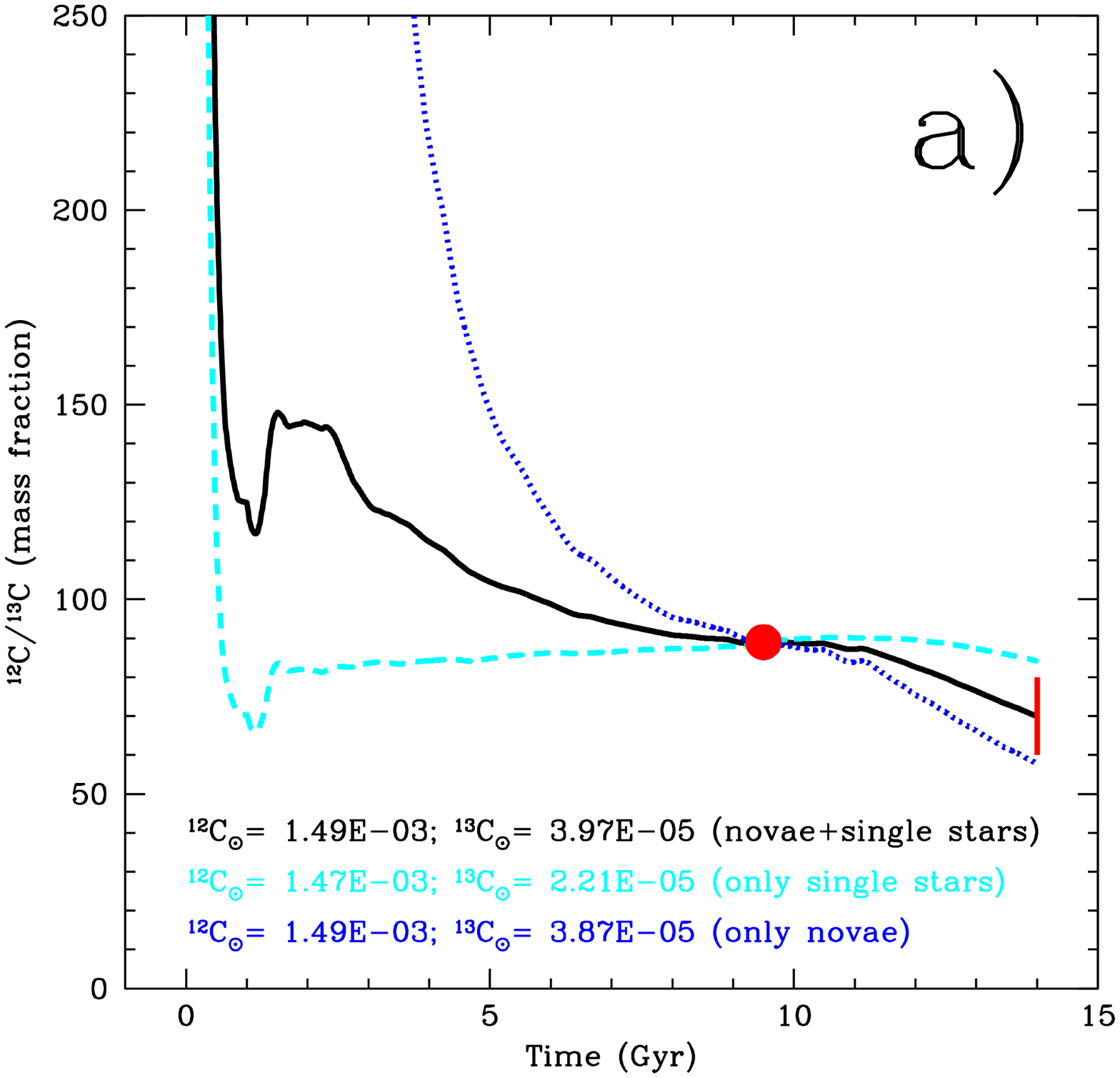,width=8.6cm,height=8.6cm} 
	    \psfig{figure=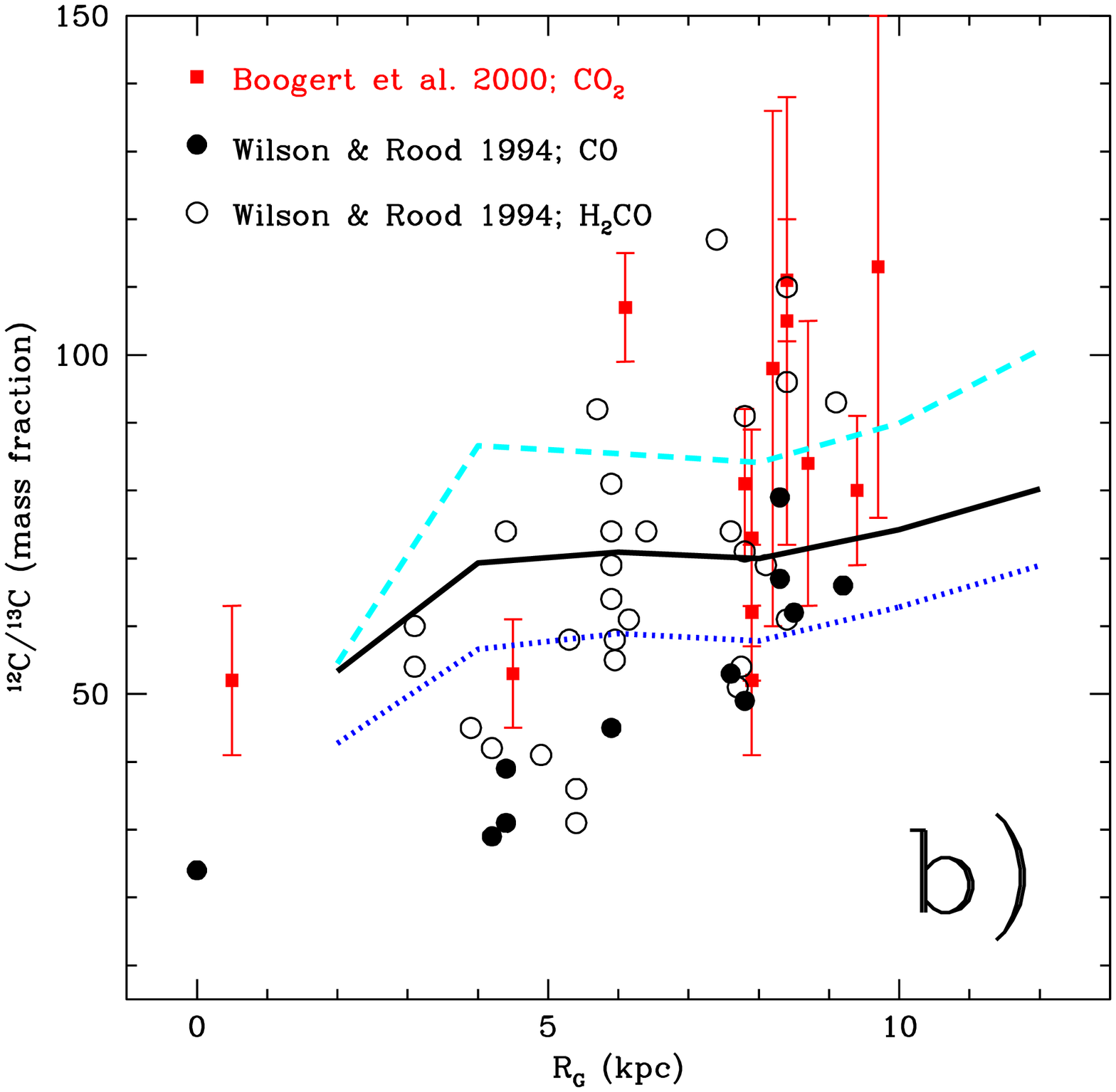,width=8.6cm,height=8.6cm}}
\caption{ {\bf a)} Temporal evolution of the carbon isotopic ratio in the 
	  solar neighbourhood in the case of $^{13}$C production from single 
	  stars alone (Model 1, {\it dashed line}), novae alone (Model 2, {\it 
          dots}) or novae plus single stars (Model 3, {\it continuous line}). 
          The model $^{12}$C/$^{13}$C ratios 4.5 Gyr ago are normalised to 
	  their solar values. The solar abundances by mass of $^{12}$C and 
	  $^{13}$C predicted by the models are given at the bottom of the 
	  figure. The big dot is the observed meteoritic ratio; the vertical 
	  bar is the ratio observed in the local ISM (references can be found 
	  in the text). The minimum in the curves at around 1 Gyr is due to 
	  the transition between halo-thick disc and thin disc in the 
	  two-infall model. 
	  {\bf b)} Theoretical present-day carbon isotopic ratio across the 
	  Galactic disc in the case of $^{13}$C production from single stars 
	  alone (Model 1, {\it dashed line}), novae alone (Model 2, {\it 
	  dots}) or novae plus single stars (Model 3, {\it continuous line}) 
	  vs. observations.}
\end{figure*}

The nucleosynthesis from novae is taken into account in the chemical evolution 
code as follows. We consider that 30\% of novae arise from systems containing 
ONe WDs and 70\% from systems containing CO WDs (e.g., Gehrz et al. 1998) and 
compute the abundance (in mass fraction) of element $i$ in the ejecta of the 
typical nova as:
\begin{equation}
\langle X_i \rangle = 0.3 \times \langle X_i \rangle_{\mathrm ONe\,nova} + 0.7 
\times \langle X_i \rangle_{\mathrm CO\,nova},
\end{equation}
where $\langle X_i \rangle_{\mathrm ONe\,nova}$ and $\langle X_i 
\rangle_{\mathrm CO\,nova}$ are the average abundances of the element $i$ in 
the ejecta of ONe novae and CO novae, respectively (Romano et al. 1999). These 
quantities are obtained in turn by averaging the results of two grids of 7 
evolutionary sequences each computed by Jos\'e \& Hernanz (1998) for ONe and 
CO WDs, respectively (see their tables~3 and 4). Once the elemental abundances 
in the ejecta of the typical nova have been estimated, we still need to assume 
a fiducial value for the mass ejected during each nova outburst. This is a 
highly uncertain quantity. The observationally derived mass distribution of 
nova ejecta indicates that the data peak close to a few 10$^{-4} M_\odot$ 
(Della Valle 2000 and references therein), but this estimate could be 
seriously challenged if the filling factor of the ejected shells is closer to 
10$^{-2}$\,--\,10$^{-5}$ rather than $\sim$ 0.1\,--\,1 (e.g., Shara et al. 
1997). We assume $M^{\mathrm ej} \sim$ 2 $\times$ 10$^{-5}$ $M_\odot$, in 
agreement with the predictions of the mass ejected during a nova outburst from 
nova hydrodynamical models, and compute the mass of the element $i$ ejected 
during the whole nova evolution as:
\begin{equation}
M^{\mathrm ej}_{\mathit i} = \langle X_i \rangle \times M^{\mathrm ej} \times 
n,
\end{equation}
where $n$ = 10$^{\,4}$ is the number of nova outbursts suffered on an average 
by each nova system.

In columns 3 and 4 of Table~1 we list the nucleosynthesis prescriptions for 
$^{12}$C, $^{13}$C, $^{14}$N, $^{15}$N, $^{16}$O and $^{17}$O from LIMS and 
massive stars adopted in our different models. Models are labelled as given in 
column 1. The average masses of $^{12}$C, $^{13}$C, $^{14}$N, $^{15}$N, 
$^{16}$O and $^{17}$O ejected by each nova system during its overall evolution 
are listed in column 5. In some cases, they differ from the average values 
obtained with the procedure described above (Eq.~3), since these latter tend 
to overestimate the solar abundances of $^{13}$C, $^{15}$N and $^{17}$O, while 
in this table we want to list the contributions we actually used in order to 
reproduce the observations. For each model, column 6 indicates if the CNO 
production from novae is treated as primary or secondary. When the CNO 
production is treated as secondary, the quantities listed in Table~1, 
corresponding to accreted envelopes of solar chemical composition, have been 
scaled according to the abundances of the seed nuclei predicted by the model 
at each time.

\section{Results}

\subsection{$^{\bmath{13}}$C production from LIMS}

It is well known that LIMS produce $^{13}$C, both as a primary and secondary 
element. $^{13}$C is also produced by massive stars, although in smaller 
amounts. In Fig.~1 we show as a dashed line the results of a model (Model~1, 
Table~1) in which $^{13}$C is made only by single stars. Nucleosynthesis 
prescriptions for LIMS and massive stars are from van den Hoek \& Groenewegen 
(1997 -- their standard case in which the mass loss parameter for stars on the 
AGB, $\eta_{\mathrm AGB}$, remains constant and equal to 4) and Nomoto et al. 
(1997), respectively. It is found that although single stars alone can explain 
the solar abundance of $^{13}$C (Table~2), they fail in accounting for the 
observed decrease of the $^{12}$C/$^{13}$C ratio in the solar neighbourhood in 
the last 4.5 Gyr (Fig.~1, {\it left panel}). Conversely, they predict a 
roughly constant $^{12}$C/$^{13}$C ratio in the past $\sim$~12 Gyr. The same 
conclusions still hold if we adopt the van den Hoek \& Groenewegen (1997) 
yields computed with $\eta_{\mathrm AGB} = 1$ for $Z = 0.001$ and 
$\eta_{\mathrm AGB} = 2$ for $Z = 0.004$ for LIMS, and the Woosley \& Weaver 
(1995) yields rather than the Nomoto et al. (1997) ones for massive stars. 
This is due to the fact that in all these models the bulk of $^{13}$C comes 
from intermediate-mass stars suffering HBB ($m \geq$ 4 $M_\odot$, $\tau_m 
\leq$ 180 Myr). Therefore, changing the yields either in massive stars or in 
the low-metallicity range does not affect our main conclusions. 
\begin{table*}
\begin{minipage}{122mm}
\caption{Predicted and observed abundances by mass at the time of Sun 
         formation ($t_\odot$ = 9.5 Gyr).}
\begin{tabular}{c c c c c c}
\hline
$^{12}$C & $^{13}$C & $^{14}$N & $^{15}$N & $^{16}$O & $^{17}$O \\
\hline
\multicolumn{6}{c}{Model 1}\\
1.47 $\times$ 10$^{-3}$ & 2.21 $\times$ 10$^{-5}$ & -- & -- &
6.92 $\times$ 10$^{-3}$ & 3.87 $\times$ 10$^{-6}$ \\
\multicolumn{6}{c}{ }\\
\multicolumn{6}{c}{Model 2}\\
1.49 $\times$ 10$^{-3}$ & 3.87 $\times$ 10$^{-5}$ & 0.96 $\times$ 10$^{-3}$ & 
3.87 $\times$ 10$^{-6}$ & 7.72 $\times$ 10$^{-3}$ & 3.89 $\times$ 10$^{-6}$ \\
\multicolumn{6}{c}{ }\\
\multicolumn{6}{c}{Model 2{\it s}}\\
1.49 $\times$ 10$^{-3}$ & 1.76 $\times$ 10$^{-5}$ & 0.95 $\times$ 10$^{-3}$ & 
3.87 $\times$ 10$^{-6}$ & 7.71 $\times$ 10$^{-3}$ & 3.85 $\times$ 10$^{-6}$ \\
\multicolumn{6}{c}{ }\\
\multicolumn{6}{c}{Model 3}\\
1.49 $\times$ 10$^{-3}$ & 3.97 $\times$ 10$^{-5}$ & 0.95 $\times$ 10$^{-3}$ & 
3.88 $\times$ 10$^{-6}$ & 7.71 $\times$ 10$^{-3}$ & 4.00 $\times$ 10$^{-6}$ \\
\multicolumn{6}{c}{ }\\
\multicolumn{6}{c}{Model 3{\it s}}\\
1.49 $\times$ 10$^{-3}$ & 3.97 $\times$ 10$^{-5}$ & 0.95 $\times$ 10$^{-3}$ & 
3.87 $\times$ 10$^{-6}$ & 7.71 $\times$ 10$^{-3}$ & 3.99 $\times$ 10$^{-6}$ \\
\multicolumn{6}{c}{ }\\
\multicolumn{6}{c}{Model 3{\it n}}\\
1.86 $\times$ 10$^{-3}$ & 4.67 $\times$ 10$^{-5}$ & 1.00 $\times$ 10$^{-3}$ & 
4.99 $\times$ 10$^{-6}$ & 6.12 $\times$ 10$^{-3}$ & 4.41 $\times$ 10$^{-6}$ \\
\multicolumn{6}{c}{ }\\
\multicolumn{6}{c}{Model 3{\it m}}\\
1.86 $\times$ 10$^{-3}$ & 4.67 $\times$ 10$^{-5}$ & 1.00 $\times$ 10$^{-3}$ & 
4.99 $\times$ 10$^{-6}$ & 6.12 $\times$ 10$^{-3}$ & 4.44 $\times$ 10$^{-6}$ \\
\multicolumn{6}{c}{ }\\
\multicolumn{6}{c}{Observed -- Anders \& Grevesse (1989)}\\
3.03 $\times$ 10$^{-3}$ & 3.65 $\times$ 10$^{-5}$ & 1.11 $\times$ 10$^{-3}$ & 
4.36 $\times$ 10$^{-6}$ & 9.59 $\times$ 10$^{-3}$ & 3.89 $\times$ 10$^{-6}$ \\
\multicolumn{6}{c}{ }\\
\multicolumn{6}{c}{Observed -- Holweger (2001)}\\
3.31 $\times$ 10$^{-3}$ & -- & 0.85 $\times$ 10$^{-3}$ & -- & 
6.24 $\times$ 10$^{-3}$ & -- \\
\multicolumn{6}{c}{ }\\
\multicolumn{6}{c}{Observed -- Allende Prieto et al. (2001, 2002)}\\
2.09 $\times$ 10$^{-3}$ & -- & -- & -- & 5.56 $\times$ 10$^{-3}$ & -- \\
\hline
\end{tabular}
\end{minipage}
\end{table*}

During HBB, $^{13}$C is mostly produced as a primary element. However, a 
secondary component is present as well, depending on how much of the 
dredged-up $^{12}$C is newly produced inside the star and how much was already 
present at the time of the star birth (see van den Hoek \& Groenewegen 1997). 
The constancy of the $^{12}$C/$^{13}$C ratio we predict in the last $\sim$~12 
Gyr is mainly due to the fact that most of the $^{13}$C is of primary origin, 
and to the fact that in the framework of our model $^{13}$C production 
proceeds almost in step with that of $^{12}$C. In order to solve the 
discrepancy with the observations, one may argue whether the $^{13}$C yields 
from single low-mass stars, which produce $^{13}$C always as a secondary 
element, are underestimated.

Indeed, in the last few years a great deal of work has appeared in the 
literature on whether standard stellar nucleosynthesis models are 
underestimating the $^{13}$C content in the material ejected from LIMS. It has 
often been suggested that the progenitors of the PNe which are observed to 
have carbon isotopic ratios below the values expected from standard AGB models 
must have undergone some non-standard mixing process, perhaps induced by 
rotation, during their red giant phase and/or AGB phase (Bachiller et al. 
1997; Palla et al. 2000; Balser, McMullin \& Wilson 2002). The existence of 
this non-standard mixing mechanism is also invoked in order to reconcile 
predictions from standard Galactic evolutionary models with observations of 
$^3$He (e.g., Galli et al. 1997; Charbonnel \& do Nascimento 1998; Chiappini, 
Renda \& Matteucci 2002).

However, the extent of the $^{13}$C enhancement as a function of mass with 
varying metallicity is, to date, rather uncertain. In fact, the requirement of 
sampling low-mass objects has often led to concentrating the observational 
efforts on metal-poor clusters. Charbonnel, Brown \& Wallerstein (1998) found 
that field giants belonging to the old Galactic disc population ([Fe/H] $\sim$ 
$-$0.6) can reach $^{12}$C/$^{13}$C $\sim$ 7. For more metal-rich stars, a 
reasonable estimate might be $\sim$ 10\,--\,15 (from observations of the 
carbon isotopic ratio in stars ascending the RGB in the open cluster M\,67; 
Gilroy \& Brown 1991). Measurements of $^{12}$C and $^{13}$C abundances in 
eleven bright giant members of the globular cluster $\omega$ Centauri spanning 
the metallicity range $-$1.8 $<$ [Fe/H] $<$ $-$0.8 do not reveal any real 
trend of the efficacy of the first giant branch mixing with metallicity 
(Smith, Terndrup \& Suntzeff 2002). The mean value for this sample is 
$\langle ^{12}$C/$^{13}$C$\rangle = 4.3 \pm 0.4$, with nine stars sharing, 
within the errors, the equilibrium ratio of 3.5. Standard stellar evolution 
theory predicts a decrease of $^{12}$C/$^{13}$C from $\sim$ 90 to $\sim$ 25 
during the first dredge-up on the RGB for LIMS (Iben 1964; Iben \& Renzini 
1984). Therefore, an extra-mixing which further processes material during the 
RGB seems to be at work.
\begin{figure}
\centerline{\psfig{figure=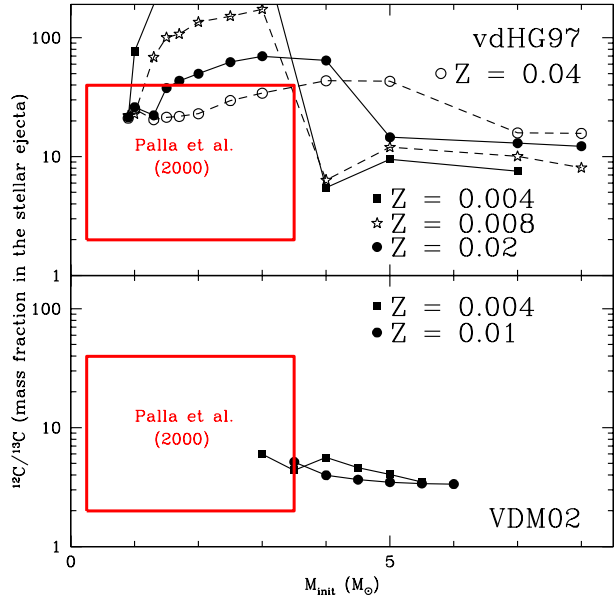,width=8.6cm,height=8.6cm}}
\caption{ $^{12}$C/$^{13}$C vs. progenitor mass in the total ejecta of LIMS.
	  Stellar models are both from van den Hoek \& Groenewegen (1997) {\it 
	  (upper panel)} and Ventura et al. (2002) {\it (lower panel)}, and 
	  have been computed for different values of the initial metallicity 
          of the stars. The box represents the region occupied by the PNe 
          detected in CO by Palla et al. (2000). Ventura et al. (2002) predict 
	  sistematically lower ratios than van den Hoek \& Groenewegen (1997) 
	  in the mass range 3\,--\,6 $M_\odot$. However, their computations do 
	  not extend to the interesting 1\,--\,3 $M_\odot$ range, where model 
	  predictions could be usefully compared to observations.}
\end{figure}
\begin{figure}
\centerline{\psfig{figure=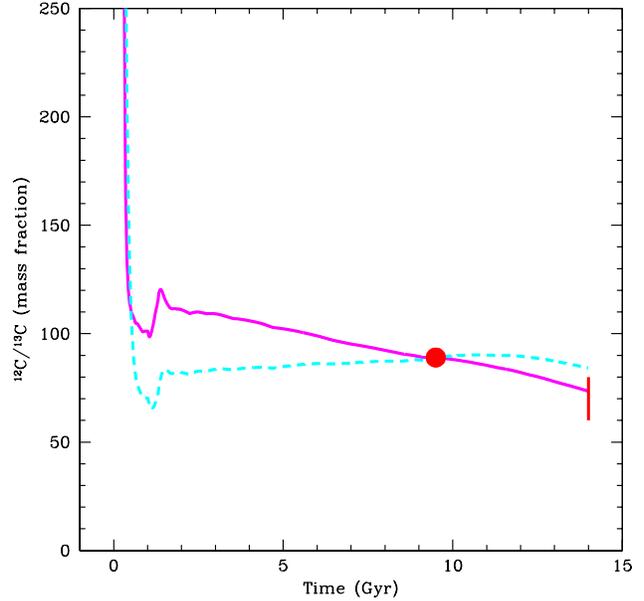,width=8.6cm,height=8.6cm}}
\caption{ In this figure we show the effect of lowering the extent of HBB
          in intermediate-mass stars on the predicted $^{12}$C/$^{13}$C ratio. 
	  The {\it dashed line} shows the results of a model that adopts the 
	  standard yields of van den Hoek \& Groenewegen (1997) (Model 1); the 
	  {\it continuous line} shows the results of a model that accounts for 
	  the minimum HBB effect still compatible with the observations.}
\end{figure}

The $^{12}$C/$^{13}$C ratio in Galactic carbon stars has been observed by 
Lambert et al. (1986), Onhaka \& Tsuji (1996, 1999), Abia \& Isern (1997) and 
Sch\"oier \& Olofsson (2000). In particular, Onhaka \& Tsuji (1996, 1999) 
found $^{12}$C/$^{13}$C ratios about a factor of 2 to 3 smaller than those of 
Lambert et al. (1986) for the same stars\footnote{ The values for three of the 
stars have been revised upwards later on by Onhaka et al. (2000) and found to 
be closer to those estimated by Lambert et al. (1986).}, whereas Abia \& Isern 
(1997) found something falling in between the results obtained by Lambert et 
al. (1986) and Onhaka \& Tsuji (1996). To solve the controversy, Sch\"oier \& 
Olofsson (2000) performed independent estimates of the $^{12}$C/$^{13}$C 
ratio, using CO radio line emission from the circumstellar envelopes, for a 
sample of carbon stars showing large discrepancies between the sets of 
photospheric estimates. They used a non-LTE radiative transfer code. The 
$^{12}$C/$^{13}$C ratios range from 2.5 to 90 and agree with those estimated 
in the photospheres by Lambert et al. (1986). Finally, in a recent paper, Abia 
et al. (2001) suggest the activation of slow circulation mechanisms below the 
convective border of the envelope for {\it most} of the observed stars, in 
order to explain the low observed $^{12}$C/$^{13}$C values. In conclusion, it 
appears that at least a fraction of PN progenitors do have reduced 
$^{12}$C/$^{13}$C ratios compared to standard model expectations, and that 
some extra-mixing should be required.

In Fig.~2 we show the $^{12}$C/$^{13}$C ratio vs. progenitor mass in the total 
ejecta of LIMS compared to observations in PNe (Palla et al. 2000). Stellar 
models are either from van den Hoek \& Groenewegen (1997) {\it (upper panel)} 
or Ventura et al. (2002) {\it (lower panel)}, and have been computed for 
different values of the initial stellar metallicity. Ventura et al. (2002) 
predict sistematically lower ratios than van den Hoek \& Groenewegen (1997) in 
the mass range 3\,--\,6 $M_\odot$. However, their computations do not extend 
to the 1\,--\,3 $M_\odot$ range, where model predictions could be usefully 
compared to observations. This is the reason why we do not adopt the carbon 
yields of Ventura et al. (2002). Notice that in this figure we are comparing 
theoretical {\it total} yields with $^{12}$C/$^{13}$C ratios observed in PNe. 
It would be better to compare the observations with the yields {\it at the tip 
of the AGB}. We choose the total yields rather than those at the end of the 
AGB phase in order to compare the results from van den Hoek \& Groenewegen 
(1997) to those from Ventura et al. (2002). In fact, these latter authors do 
not give the yields at the tip of the AGB, but only the total ones. 
Theoretical $^{12}$C/$^{13}$C ratios at the tip of the AGB from van den Hoek 
\& Groenewegen (1997) are in general higher than those in the total ejecta. It 
is immediately seen that the adopted yields overestimate the $^{12}$C/$^{13}$C 
ratio in the ejecta of low-mass stars. However, as demonstrated by Palla et 
al. (2000), as far as $^{12}$C/$^{13}$C is concerned, the fraction of low-mass 
stars experiencing deep mixing does not affect significantly the overall 
results of the chemical evolution models, due to the fact that the Galactic 
evolution of $^{12}$C and $^{13}$C is mainly governed by stars in which this 
process is not expected to occur. Therefore, the main conclusions reached in 
the present paper should not be largely affected by the adoption of stellar 
yields taking extra-mixing into account.
\begin{figure*}
\centerline{\psfig{figure=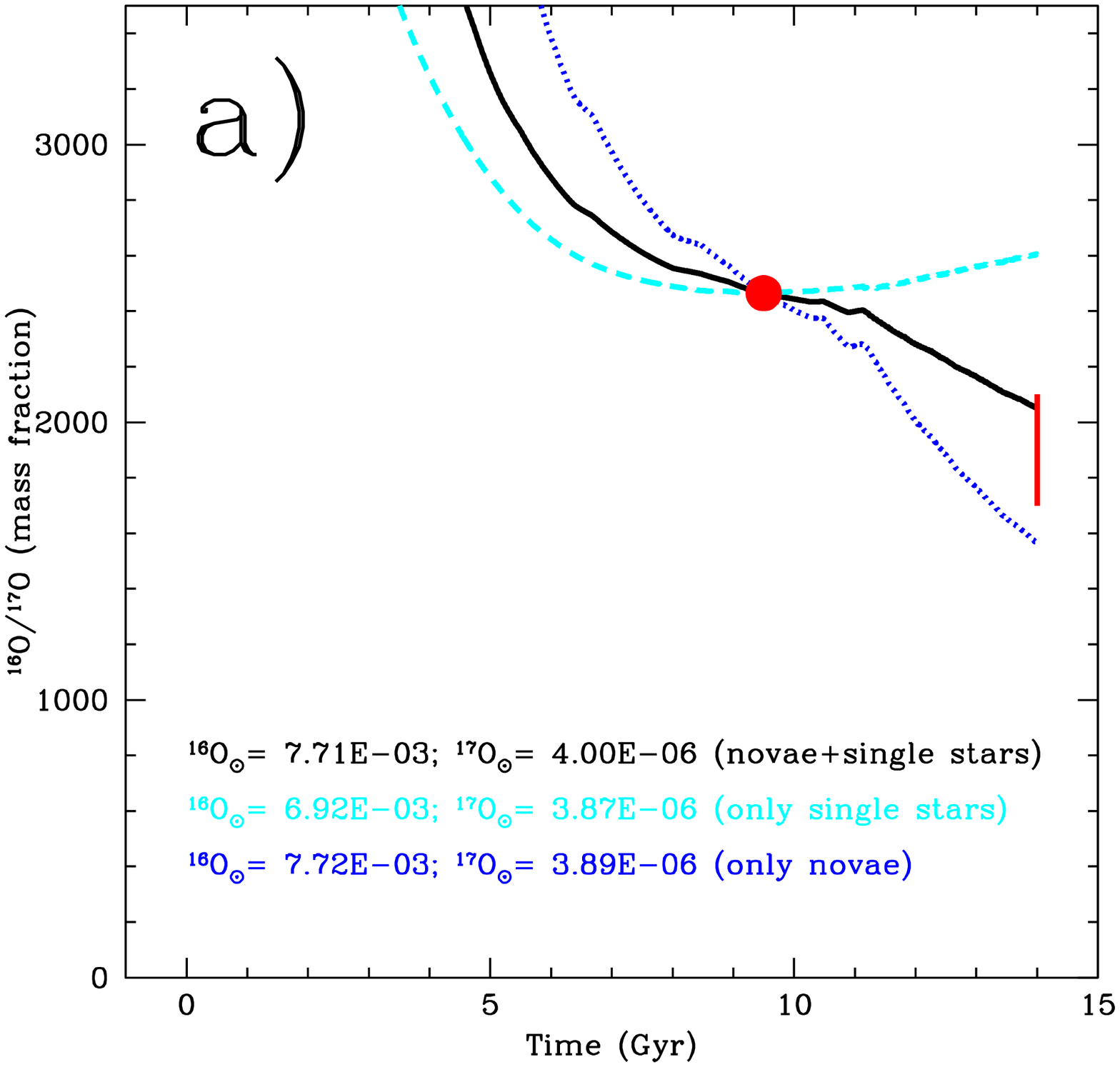,width=8.6cm,height=8.6cm}
	    \psfig{figure=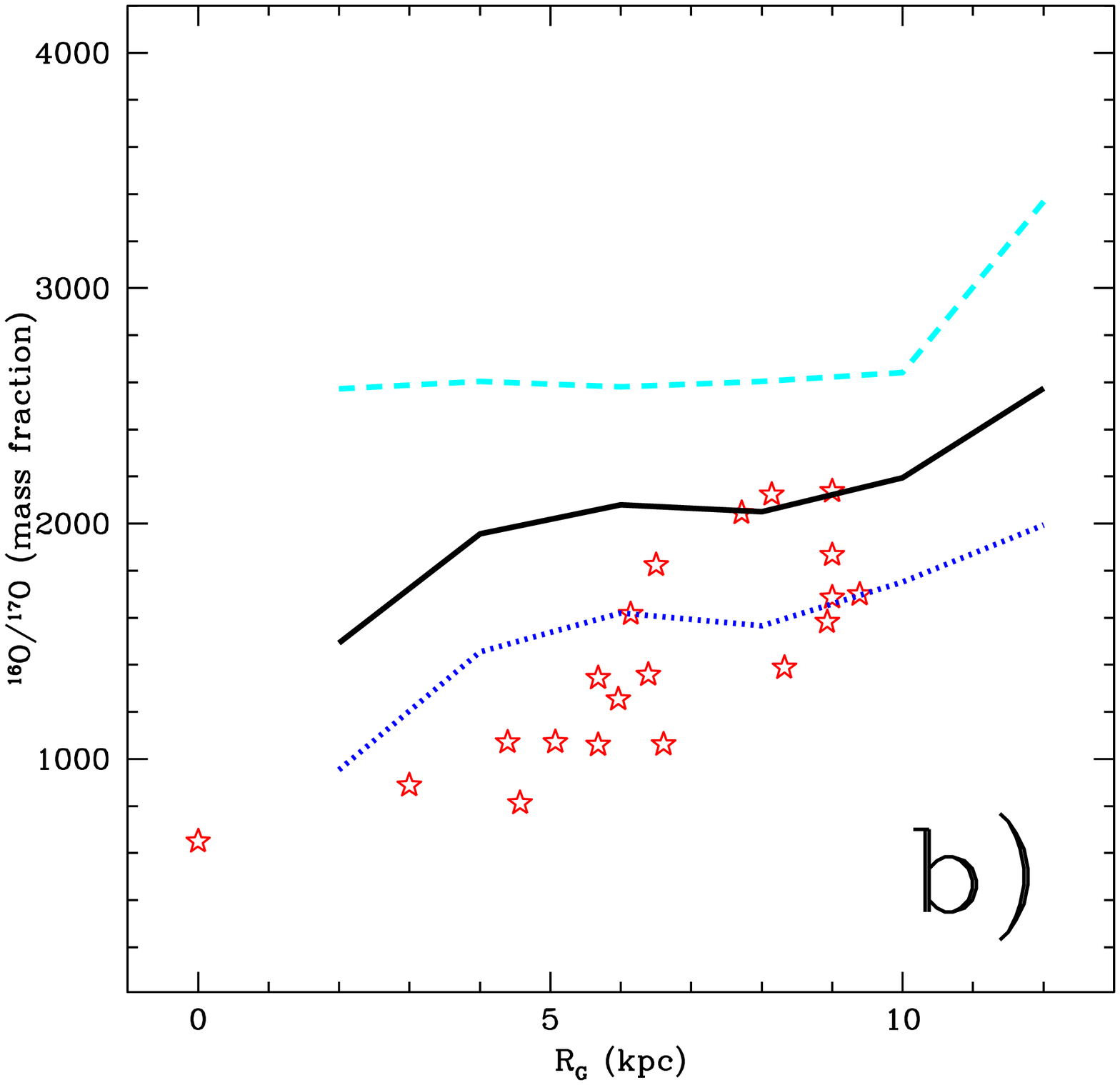,width=8.6cm,height=8.6cm}}
\caption{ {\bf a)} Temporal evolution of the oxygen isotopic ratio in the 
	  solar neighbourhood in the case of $^{17}$O production from only 
	  single stars (Model 1, {\it dashed line}), only novae (Model 2, {\it 
	  dotted line}) or novae plus single stars (Model 3, {\it continuous 
	  line}). The model $^{16}$O/$^{17}$O ratios 4.5 Gyr ago are 
	  normalised to their solar values. The solar abundances by mass of 
	  $^{16}$O and $^{17}$O predicted by the models are also given at the 
	  bottom of the figure. The big dot is the observed meteoritic ratio; 
	  the vertical bar is the ratio observed in the local ISM (see text 
	  for references).
	  {\bf b)} Theoretical present-day oxygen isotopic ratio across the 
	  Galactic disc in the case of $^{17}$O production from only single 
	  stars (Model 1, {\it dashed line}), only novae (Model 2, {\it dotted 
	  line}) or novae plus single stars (Model 3, {\it continuous line}) 
          vs. observations. Data are from Prantzos et al. (1996).}
\end{figure*}

On the other hand, the discrepancy between model predictions and observations 
can be solved if the adopted stellar yields overestimate the effect of HBB: 
running a model which adopts AGB yields computed with less HBB (tables from 22 
to 31 of van den Hoek \& Groenewegen 1997) leads to a $^{12}$C/$^{13}$C ratio 
which decreases in the last 12 Gyr (Fig.~3), in agreement with observations. 
However, in this case the predicted solar abundance of $^{13}$C is lower 
($^{13}$C$_\odot$ = 1.48 $\times$ 10$^{-5}$). Notice also that the level of 
HBB required by this model is {\it the lowest one still compatible with 
observations} in stars.

In Section~4.3 we examine in detail the possibility that novae contribute to 
the $^{13}$C production.

\subsection{$^{\bmath{17}}$O production from single stars}

$^{17}$O is produced by the NO cycle in both intermediate- and high-mass 
stars; contrary to $^{13}$C, it is not produced in the 1\,--\,2 $M_\odot$ mass 
range where the NO cycle does not operate. It is mostly produced as a 
secondary element, owing to the dependence of its abundance on the initial 
$^{16}$O content of the star. In Figs.~4a,b we show the predictions from 
Model~1 ($^{17}$O production from single stars alone) as {\it dashed lines}. 
The nucleosynthesis prescriptions for $^{17}$O from LIMS and massive stars are 
from Ventura et al. (2002) and Nomoto et al. (1997), respectively (see 
Table~1). It is worth noticing that, since the study of Nomoto et al. (1997) 
concerns only the He-cores and does not treat properly the products of the 
H-burning surviving outside the receding He-core, the yields of $^{17}$O from 
massive stars adopted by Model~1 are underestimated. Therefore, in this model 
$^{17}$O comes practically solely from intermediate-mass stars in the mass 
range 2.5\,--\,6 $M_\odot$. Nevertheless, in this model we have to reduce the 
$^{17}$O yields by Ventura et al. (2002) by a factor of $\sim$ 6 in order not 
to overestimate the $^{17}$O solar abundance. In fact, Ventura et al. (2002) 
predict a too large production factor for $^{17}$O. This problem was already 
encountered by Prantzos et al. (1996), who adopted the $^{17}$O yields from 
LIMS by Marigo et al. (1996). They showed that the Marigo et al. yields have 
to be reduced by a factor of $\sim$ 3 in order not to overproduce the solar 
$^{17}$O (cfr. figure~1f of Prantzos et al. 1996 and figure~7 of Ventura et 
al. 2002).

Model~1 predicts that the $^{16}$O/$^{17}$O ratio decreases during the first 
$\sim$ 7 Gyrs of Galactic evolution, then it flattens out and finally 
increases at around $t$ = 10 Gyr. At first glance, this is unexpected, since 
the $^{16}$O/$^{17}$O ratio is a typical primary to secondary ratio and hence, 
according to common wisdom, it should always decrease in time. The explanation 
of this behavior stems from the stellar mass range from which $^{17}$O 
originates in our model, and from the star formation history. The yields of 
Ventura et al. (2002) we adopt here are computed for stars in the mass range 
2.5\,--\,6 $M_\odot$. The lifetime of a 6 $M_\odot$ star is $\sim$ 0.065 Gyr, 
while that of a 2.5 $M_\odot$ star is $\sim$ 0.75 Gyr. Therefore, stars in 
this mass range start restoring their newly synthesized $^{17}$O rather early 
during Galaxy's evolution. An age of 7 Gyr (the age at which the oxygen 
isotopic ratio flattens out) corresponds to the lifetime of a 1.4 $M_\odot$ 
star, while an age of 10 Gyr (the age at which the oxygen isotopic ratio 
starts increasing) corresponds to the lifetime of a 1 $M_\odot$ star. Stars 
with $m <$ 2.5 $M_\odot$ do not produce $^{17}$O in our model. From this fact 
and from an inspection of Fig.~5, it can be immediately understood why, 
although $^{17}$O is produced as a secondary element, the $^{16}$O/$^{17}$O 
ratio first flattens and then increases in time: this happens because a 
significant fraction of the stars which die at those times are not $^{17}$O 
producers. Moreover, they formed out of gas characterized by high 
$^{16}$O/$^{17}$O ratios, and so the fraction of unburnt material that they 
restore into the ISM is characterized by the same high ratios. The predicted 
gradient across the Galactic disc can be easily explained by the same 
arguments. In particular, the high $^{16}$O/$^{17}$O ratio we find at inner 
radii at the present time is due to the formation of an important fraction of 
low-mass stars at early times, out of gas with a high $^{16}$O/$^{17}$O 
content, which die at late times and restore into the ISM matter characterized 
by this high $^{16}$O/$^{17}$O ratio -- and low metal abundance (Fig.~5, {\it 
top panel}).
\begin{figure*}
\centerline{\psfig{figure=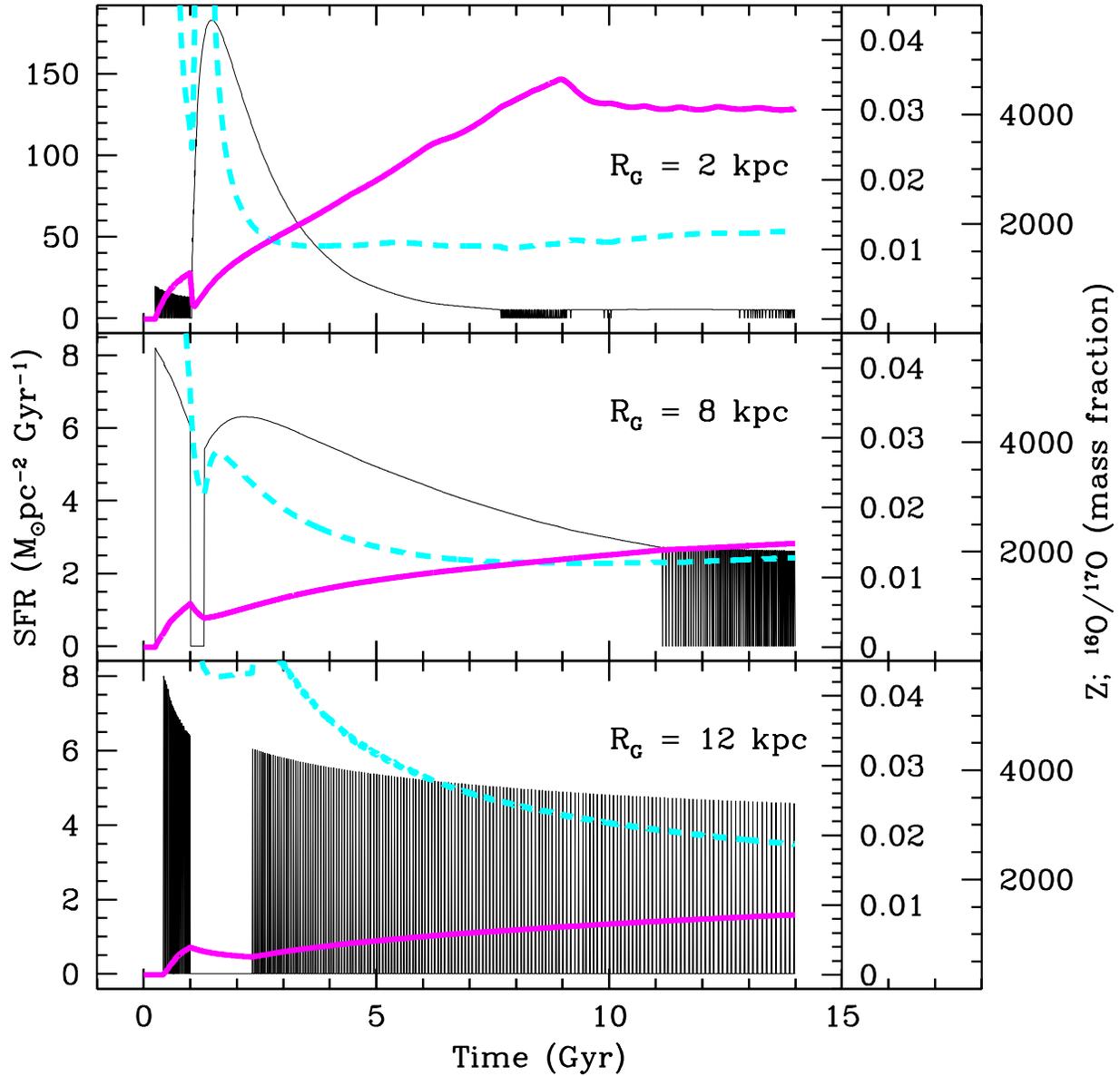,width=17.2cm,height=17.2cm}}
\caption{ The predicted temporal evolution of the total metal content ($Z$) 
	  and of the $^{16}$O/$^{17}$O ratio in the ISM for three different 
	  Galactocentric distances are shown as a thick solid and a thick 
	  dashed line, respectively. The star formation rate as a function of 
	  time is also shown, as a thin solid line. Its oscillating behavior 
	  at large Galactocentric distances is due to the presence of a 
	  threshold in the gas density below which the star formation stops 
	  (see Chiappini et al. 1997).}
\end{figure*}
\begin{table}
\caption{ Predicted and observed gradients (dex kpc$^{-1}$) at the present 
	  time ($t_{\mathrm{Gal}}$ = 14 Gyr).}
\begin{tabular}{@{} c c c @{}}
\hline
$\Delta$($^{12}$C/$^{13}$C)/$\Delta$R$_G$ & 
$\Delta$($^{14}$N/$^{15}$N)/$\Delta$R$_G$ & 
$\Delta$($^{16}$O/$^{17}$O)/$\Delta$R$_G$ \\
\hline
\multicolumn{3}{c}{Model 1}\\
4.6 & -- & 79 \\
\multicolumn{3}{c}{ }\\
\multicolumn{3}{c}{Model 2}\\
2.6 & 8 & 104 \\
\multicolumn{3}{c}{ }\\
\multicolumn{3}{c}{Model 2{\it s}}\\
12 & 26 & 255 \\
\multicolumn{3}{c}{ }\\
\multicolumn{3}{c}{Model 3}\\
2.7 & 8 & 108 \\
\multicolumn{3}{c}{ }\\
\multicolumn{3}{c}{Model 3{\it s}}\\
4.3 & 26 & 226 \\
\multicolumn{3}{c}{ }\\
\multicolumn{3}{c}{Model 3{\it n}}\\
4.8 & 19 & 164 \\
\multicolumn{3}{c}{ }\\
\multicolumn{3}{c}{Model 3{\it m}}\\
4.8 & 18 & 193 \\
\multicolumn{3}{c}{ }\\
\multicolumn{3}{c}{Observed}\\
4.5 $\pm$ 2.2$^{\mathrm a}$ & 19.7 $\pm$ 8.9$^{\mathrm b}$ & -- \\
\hline
\end{tabular}

$^{\mathrm a}${Boogert et al. (2000).}\\
$^{\mathrm b}${Dahmen et al. (1995).}
\end{table}

Summarizing: stars in the mass range 2.5\,--\,6 $M_\odot$ are able to 
reproduce the solar abundance of $^{17}$O without requiring further $^{17}$O 
producers, but their theoretical yields have to be reduced (e.g., by a factor 
of $\sim$ 6 in the case of the $^{17}$O yields adopted in this work). However, 
once the solar $^{17}$O abundance is explained, theoretical predictions on 
both the temporal evolution of the $^{16}$O/$^{17}$O ratio in the solar 
neighbourhood and the $^{16}$O/$^{17}$O gradient along the Galactic disc are 
not in agreement with the observations. 

It is interesting to note that we reach the same conclusion also if we adopt 
the metallicity dependent yields by Woosley \& Weaver (1995) for massive stars 
and set them as the only $^{17}$O producers. In fact, by using the metallicity 
dependent yields by Woosley \& Weaver (1995) for massive stars, the solar 
abundance of $^{17}$O is reproduced (within the errors), but the oxygen 
isotopic ratio remains almost flat in the past 4.5 Gyr. This result disagrees 
with previous investigations suggesting that massive stars alone can explain 
the observed evolution of $^{16}$O/$^{17}$O in the Galaxy, once the 
metallicity effects on the stellar yields are properly taken into account 
(e.g., Timmes et al. 1995; Goswami \& Prantzos 2000). We argue that the 
differences could be due to the fact that in our models many stars in the mass 
range 1\,--\,2 $M_\odot$ die and restore matter with high $^{16}$O/$^{17}$O at 
late times. Moreover, the effect of the threshold in the star formation 
process can play a role in the same direction.

In conclusion, our results suggest: {\it i)} the need for a major revision of 
the theoretical $^{17}$O yields from intermediate-mass stars on the one hand 
and {\it ii)} the occurrence of late $^{17}$O pollution from some different, 
long-lived stellar source on the other.

\subsection{CNO production during nova outbursts}

In previous papers (Romano et al. 1999, 2001) we adopted the $^7$Li yields 
from novae by Jos\'e \& Hernanz (1998) and showed that novae are among the 
best candidates in order to explain the sudden increase of the lithium 
abundance in the ISM at [Fe/H] $>$ $-$0.5 dex, as traced by dwarf stars in the 
solar vicinity hotter than 5700 K in the $\log \epsilon (^7$Li$)$\,--\,[Fe/H] 
diagram. The key ingredients which led to this result were {\it i)} the high 
overproduction factors of $^7$Li with respect to its solar abundance and {\it 
ii)} the long delay with which novae restore their nuclearly processed 
material into the ISM.

Large overproduction factors have been found also for $^{13}$C, $^{15}$N and 
$^{17}$O (see figures~1 to 3 of Jos\'e \& Hernanz 1998). These overproduction 
factors are so large that one can argue that novae can eventually account for 
most of the Galactic $^{13}$C, $^{15}$N and $^{17}$O (see the discussion in 
Section~1). Moreover, it is apparent that novae, restoring their newly 
synthesized $^{13}$C, $^{15}$N and $^{17}$O to the ISM on very long 
time-scales, act in the right direction in solving any discrepancy between 
model predictions and observations when dealing with the Galactic evolution of 
the CNO isotopic ratios.

In Figs.~1, 4 and 6 we show as dotted lines the results of a detailed chemical 
evolution model in which {\it all} of the Galactic $^{13}$C, $^{15}$N and 
$^{17}$O are supposed to come from nova outbursts (Model~2, Table~1). In this 
model, the $^{13}$C, $^{15}$N and $^{17}$O production from novae is treated as 
primary, i.e., $^{13}$C, $^{15}$N and $^{17}$O are assumed to be produced 
independently of the initial metallicity in the nova outburst. The average 
masses ejected in the form of $^{12}$C, $^{13}$C, $^{14}$N, $^{15}$N, $^{16}$O 
and $^{17}$O assumed by the model are listed in column 5 of Table~1. We assume 
that the mean mass ejected during a single burst is $\sim$ 2 $\times$ 
10$^{-5}$ $M_\odot$ and that each nova experiences 10$^4$ bursts during its 
life (see Section~3.2). Nucleosynthesis prescriptions for LIMS and massive 
stars are from van den Hoek \& Groenewegen (1997 -- their {\it standard case} 
in which the mass loss parameter for stars on the AGB, $\eta_{\mathrm AGB}$, 
is equal to 4 and does not vary with metallicity) and Nomoto et al. (1997), 
respectively. In this model, the $^{13}$C, $^{15}$N and $^{17}$O production 
from LIMS and massive stars is set to zero. It is found that it is necessary 
to lower the average nova yields of $^{13}$C, $^{15}$N and $^{17}$O from 
Jos\'e \& Hernanz (1998) in order to avoid to overestimate their meteoritic 
abundances.
\begin{figure*}
\centerline{\psfig{figure=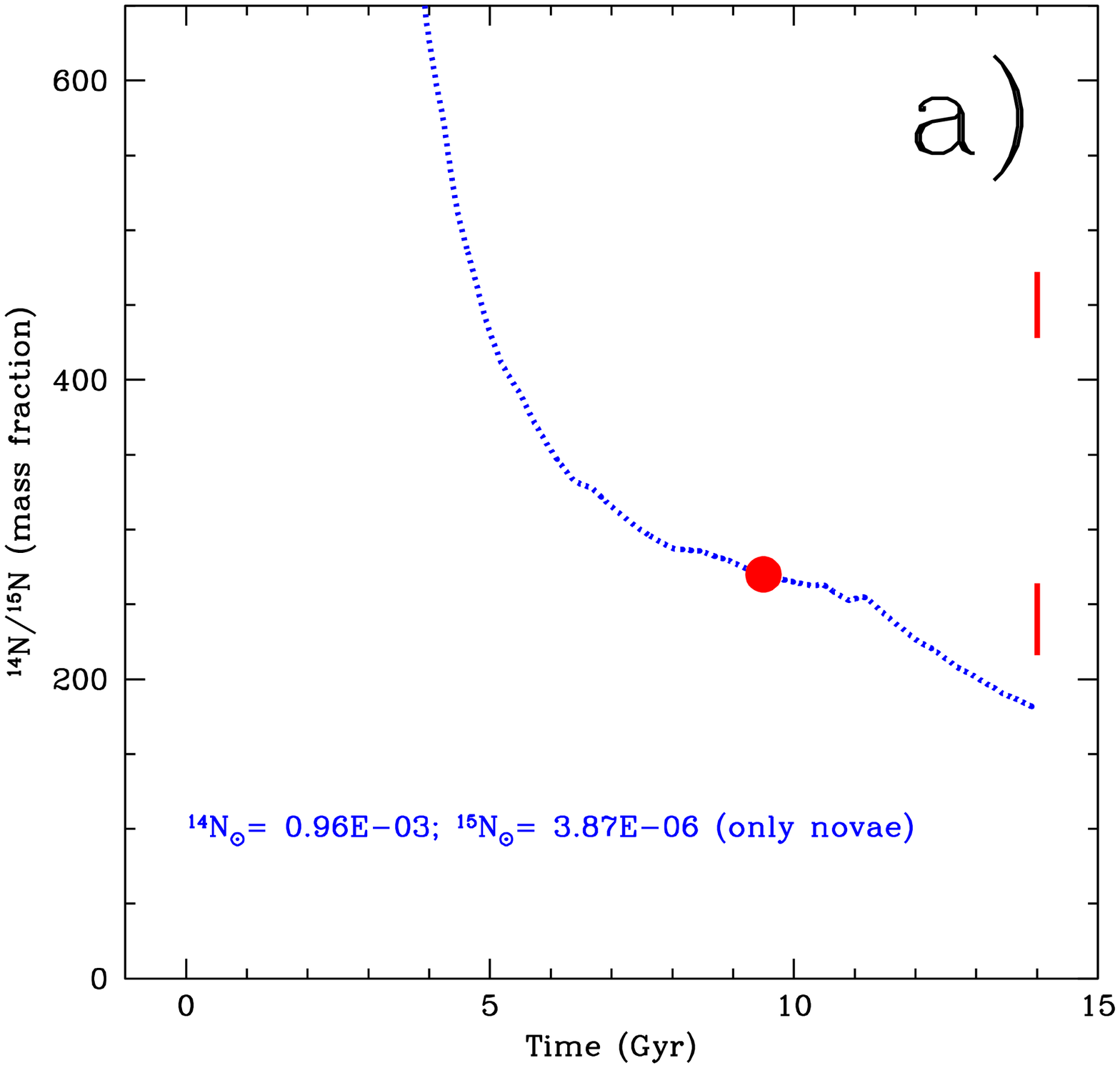,width=8.6cm,height=8.6cm}
	    \psfig{figure=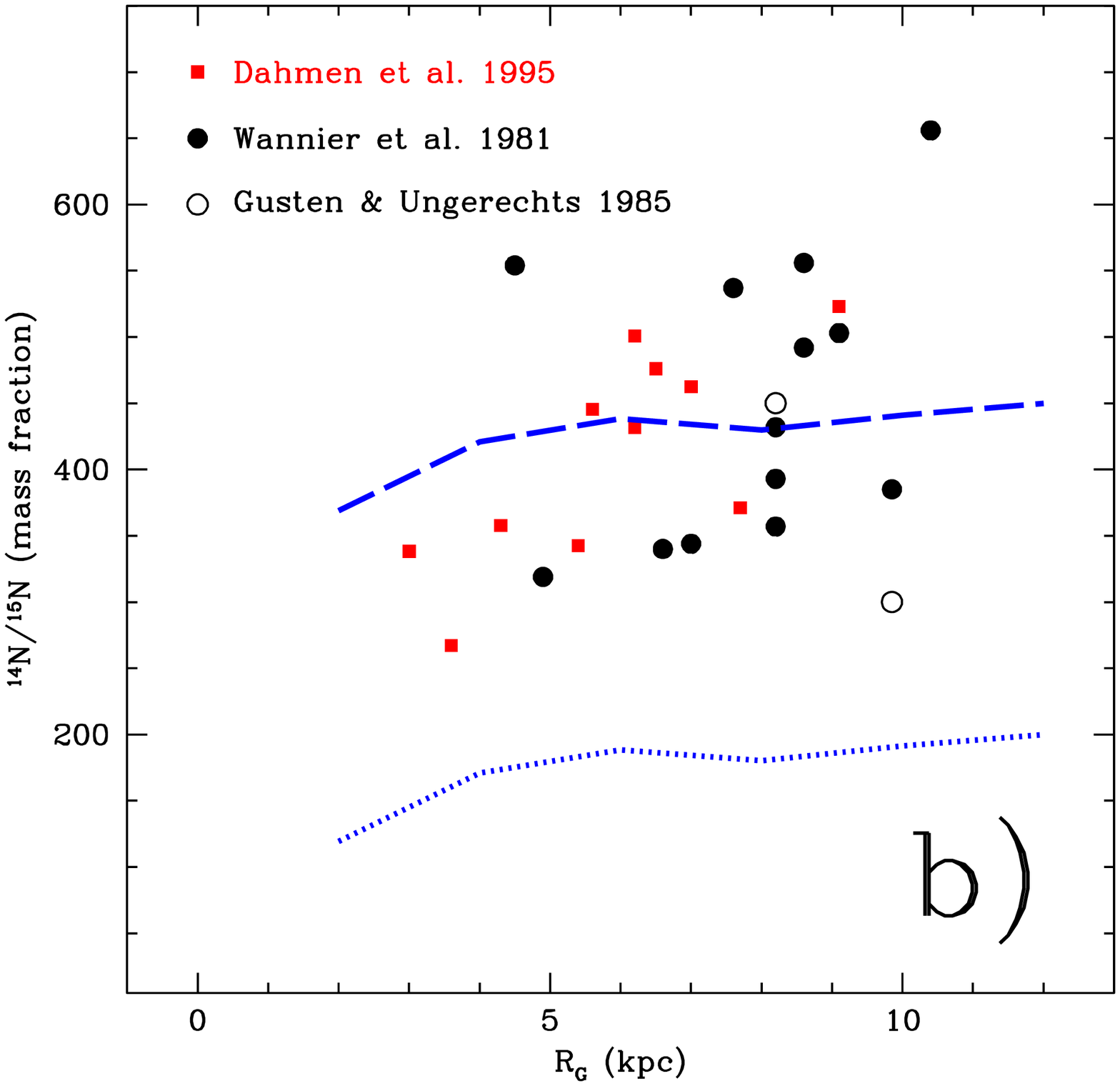,width=8.6cm,height=8.6cm}}
\caption{ {\bf a)} Temporal evolution of the nitrogen isotopic ratio in the 
	  solar neighbourhood. $^{15}$N is only produced during nova 
	  outbursts. The model $^{14}$N/$^{15}$N ratio 4.5 Gyr ago is 
	  normalised to its solar value. The solar abundances by mass of 
	  $^{14}$N and $^{15}$N predicted by the model are given at the bottom 
	  of the figure. The big dot is the observed meteoritic ratio; the 
	  vertical bars are the ratios observed in local diffuse (lower value) 
	  or dense (higher value) clouds (see text for references).
	  {\bf b)} Theoretical present-day nitrogen isotopic ratio across the 
	  Galactic disc. The {\it dashed line} are the actual model 
	  predictions {\it (dotted line)} offset to better compare the slope 
	  of the gradient with the data.}
\end{figure*}

In Figs.~1, 4 and 6, {\it panels a}, the temporal evolution of the CNO 
isotopic ratios in the solar neighbourhood is displayed. If novae are the only 
sources of $^{13}$C, $^{15}$N and $^{17}$O and produce them as {\it primary} 
elements, once the solar values are reproduced, the present-day local isotopic 
ratios fall to values smaller than observed (unless the 2\,$\sigma$ level of 
uncertainty for measurements in the local ISM is assumed), due to the delayed 
contribution of too large amounts of $^{13}$C, $^{15}$N and $^{17}$O from nova 
systems at late times. Moreover, the slope of the theoretical gradients is 
sistematically flatter than observed (Table~3).

If instead $^{13}$C, $^{15}$N and $^{17}$O from novae are {\it secondary} 
(Model~2{\it s}, Table~1), we obtain that: {\it i)} the decrease of the CNO 
isotopic ratios in the solar vicinity in the last 4.5 Gyr is more pronounced 
(Fig.~7, {\it left panels}); {\it ii)} the gradients steepen out (Fig.~7, {\it 
right panels}); in particular, in the case of nitrogen the agreement with the 
observations improves, while in the case of carbon it worsens (Table~3).
\begin{figure*}
\centerline{\psfig{figure=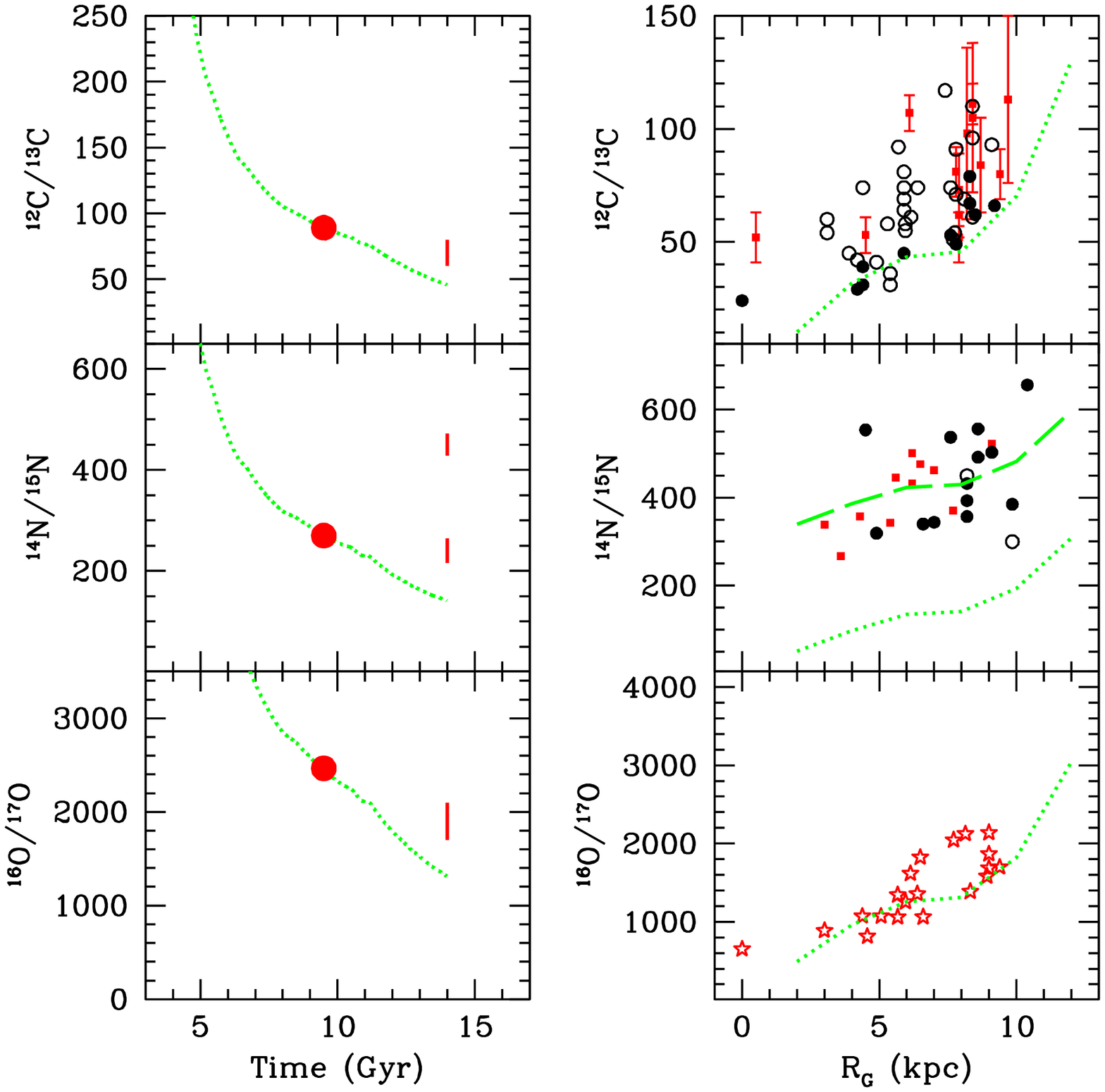,width=17.2cm,height=17.2cm}}
\caption{ The theoretical behavior of the CNO isotopic ratios in the solar 
	  neighbourhood as a function of time {\it (left panels)} and across 
	  the Galactic disc at the present time {\it (right panels)} is 
	  shown for Model~2{\it s} as a dotted line compared to observations 
	  (see text and Figs.~1, 4 and 6 for references). Model~2{\it s} is 
	  the same as Model~2, i.e., $^{13}$C, $^{15}$N and $^{17}$O are 
	  assumed to originate only from novae, but here the $^{13}$C, 
	  $^{15}$N and $^{17}$O production from novae is assumed to be 
	  secondary rather than primary (see Table~1 and text for more 
	  details). The decrease of the CNO isotopic ratios in the solar 
	  neighbourhood in the last 4.5 Gyr is inconsistent with the 
	  observations, also if the 2\,$\sigma$ level of uncertainty for 
	  measurements in the local ISM is assumed.}
\end{figure*}

It is interesting to analyse what is the effect of changing the free 
parameters involved in the computation of the nova outburst rate on the 
predicted CNO ratios. Changing the value of $\Delta t$, i.e., the delay time 
required to cool the WD at a level that guarantees strong enough nova bursts, 
from 2 to 5 Gyr, results in a lower nova rate at the present time, and hence 
lower solar abundances for the CNO group elements (both in the case in which 
the production of the minor CNO isotopes from novae is treated as primary and 
in the case in which it is treated as secondary), unless the value of $\alpha 
\, n$ is properly increased\footnote{Notice that we cannot put constraints 
upon $\alpha$ and $n$ singularly, but rather on the product $\alpha \, n$.}. 
Increasing $\alpha$ so that the value of 
$R_{\mathrm{novae}}(t_{\mathrm{Gal}})$ predicted with $\Delta t$ = 5 Gyr is 
the same as that found with $\Delta t$ = 2 Gyr results in flattening too much 
the gradients along the disc.

It is clear that the nova contribution to the chemical evolution of the 
Galaxy is at present only poorly constrained: the rate of nova outbursts in 
the past is virtually unknown; the delay time $\Delta t$ could vary from 
system to system; the mass ejected by a single nova during each burst is very 
uncertain (see D'Antona \& Matteucci 1991 and the discussion in Section~3.2). 
However, our analysis allows us to remark at least two important things: {\it 
i)} first, if the fraction of WDs which enter the formation of nova systems is 
constant in time (similarly to what is assumed for the fraction of WDs 
entering the formation of binary systems ending up as Type Ia SNe), the nova 
yields of $^{13}$C, $^{15}$N and $^{17}$O computed by Jos\'e \& Hernanz for a 
solar composition of the matter accreted by the WD overestimate the solar 
abundances of $^{15}$N and $^{17}$O; {\it ii)} second, if novae are the only 
sources of $^{13}$C, $^{15}$N and $^{17}$O, the predicted $^{12}$C/$^{13}$C, 
$^{14}$N/$^{15}$N and $^{16}$O/$^{17}$O ratios tend to decrease too fast from 
the time of Sun formation up to now, especially in the case in which their 
yields are scaled with the initial stellar metallicity.

In Figs.~1 and 4, we also show as continuous lines the predictions of a model 
(Model~3, Table~1) in which $^{13}$C and $^{17}$O are produced both by single 
stars and novae (as primary elements, in this latter case). The decrease of 
the $^{12}$C/$^{13}$C ratio in the solar neighbourhood in the last 4.5 Gyr can 
be now reproduced without requiring a reduction of the HBB strength in the van 
den Hoek \& Groenewegen (1997) standard set of yields. Also the gradient of 
this same ratio across the disc is well reproduced. However, in order not to 
overproduce the $^{17}$O abundance in the Sun, we need to lower in Model~3 the 
$^{17}$O yields from both novae and intermediate-mass stars. In particular, in 
this model novae produce almost 50 per cent of the solar $^{17}$O, while the 
remaining 50 per cent comes from intermediate-mass stars. Actually, favouring 
a higher nova contribution with respect to that from LIMS would improve the 
agreement with the observations. However, any percentage would be very 
tentative, due to the uncertainties in the nova and intermediate-mass star 
yields just mentioned before. Therefore, any acceptable solution would be by 
no means unique.
\begin{figure*}
\centerline{\psfig{figure=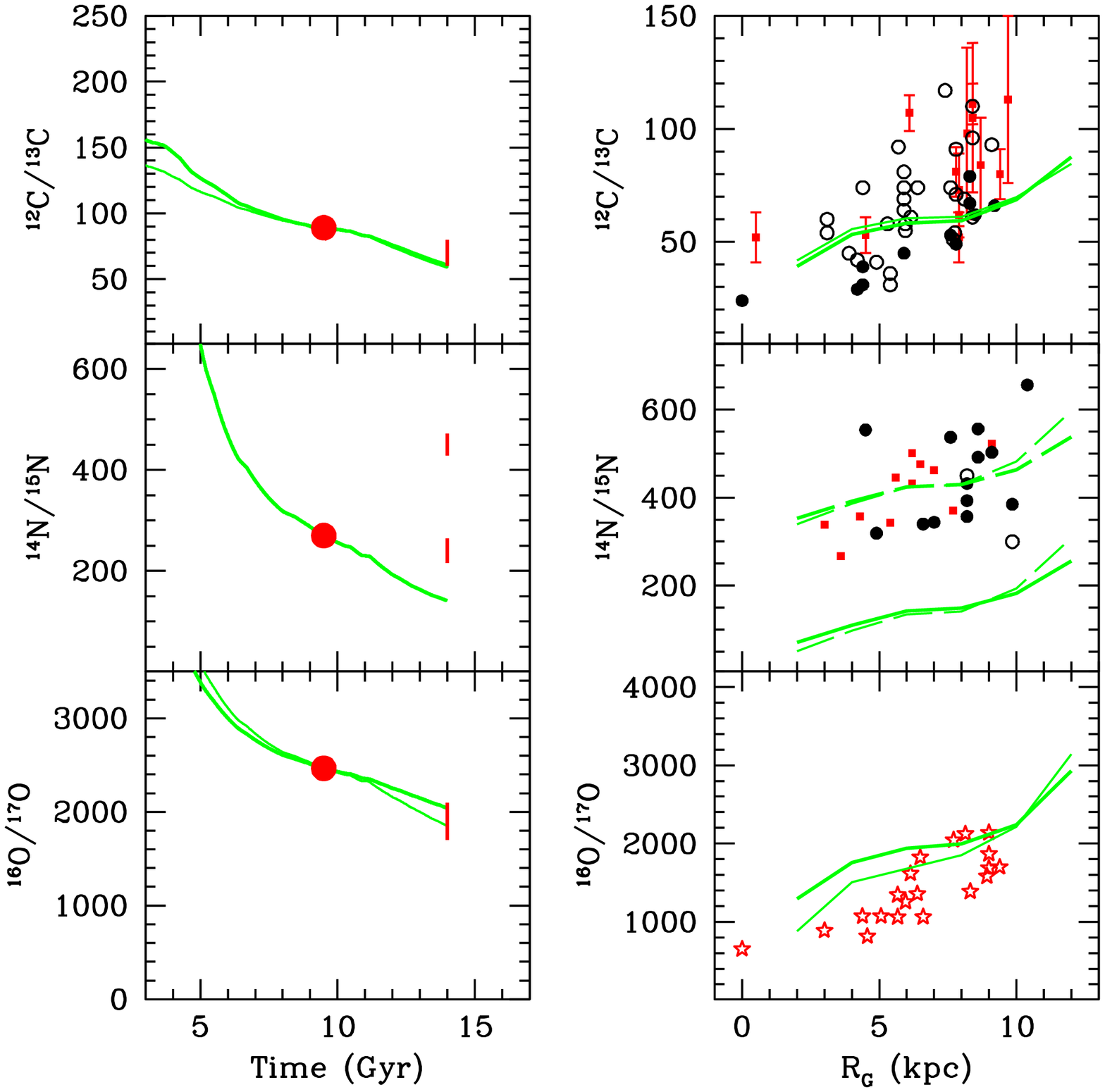,width=17.2cm,height=17.2cm}}
\caption{ The theoretical behavior of the CNO isotopic ratios in the solar 
	  neighbourhood as a function of time {\it (left panels)} and across 
	  the Galactic disc at the present time {\it (right panels)} is 
	  shown for Models~3{\it s} and 3{\it n} as a thin solid and a thick 
	  solid line, respectively, and compared to observations (see text and 
	  Figs.~1, 4 and 6 for references). As far as nova nucleosynthesis is 
	  concerned, Models~3{\it s} and 3{\it n} differ from Model 3 in the 
	  fact that the yields from novae are treated as secondary, i.e., they 
	  are scaled according to the initial metallicity of the star. 
	  Model~3{\it n} is the same as Model~3{\it s}, except for the 
	  nucleosynthesis prescriptions for single low-, intermediate- and 
	  high-mass stars (see Table~1).}
\end{figure*}
\begin{figure*}
\centerline{\psfig{figure=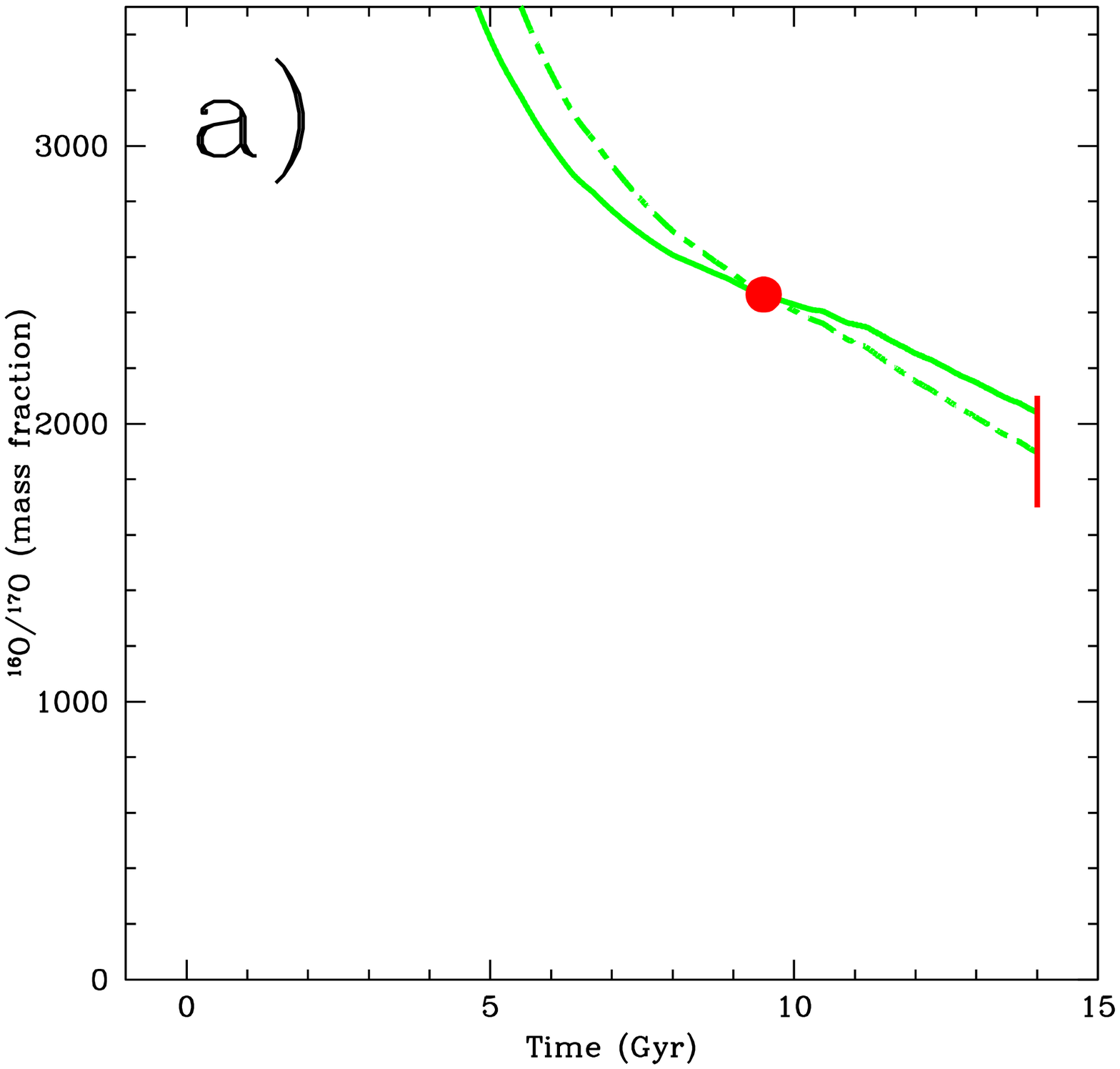,width=8.6cm,height=8.6cm}
	    \psfig{figure=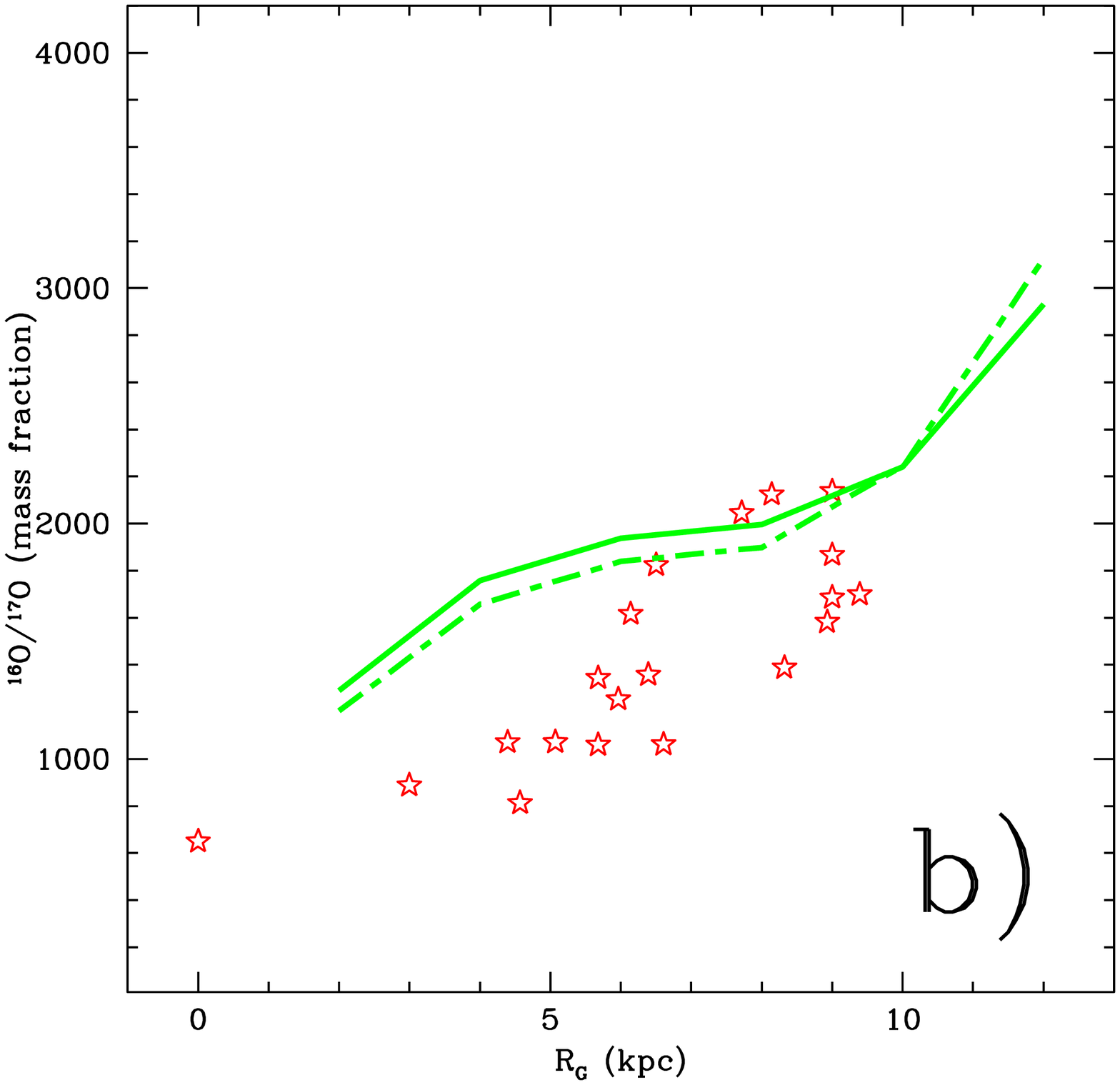,width=8.6cm,height=8.6cm}}
\caption{ {\bf a)} Temporal evolution of the oxygen isotopic ratio in the 
	  solar neighbourhood for two models in which $^{17}$O originates from 
	  intermediate- and high-mass stars as well as from novae 
	  (Models~3{\it n} and 3{\it m}, Table~1). The two models differ in 
	  the treatment of pristine $^{17}$O in stars: Model~3{\it n} {\it 
	  (continuous line)} assumes that all the $^{17}$O initially present 
	  in the gas out of which the star forms is destroyed later on by 
	  proton captures in the stellar interior, while Model~3{\it m} {\it 
	  (long-dashed-short-dashed line)} assumes that all the pristine 
	  $^{17}$O is preserved and ejected into the ISM at the death of the 
	  star.
	  {\bf b)} Theoretical oxygen isotopic ratio across the Galactic disc 
	  vs. observations for the same models as in {\it panel a}.}
\end{figure*}

\subsection{Changing the nucleosynthesis prescriptions for massive stars and 
	    LIMS}

Finally, one could ask how firm are the results we presented in the previous 
section with respect to changes in the nucleosynthesis prescriptions for 
novae, massive stars and LIMS; in other words, how much do our conclusions 
depend on the choice of the particular set of stellar yields we have adopted? 
In this section we show the results of a chemical evolution model, Model~3{\it 
s}, similar to Model~3, but where the yields from novae are scaled according 
to the initial metallicity of the star (Fig.~8, {\it thin lines}). In the same 
figure, we also show results from a model, Model~3{\it n}, similar to 
Model~3{\it s}, except for the fact that the yields for LIMS from van den Hoek 
\& Groenewegen (1997) are computed with $\eta_{\mathrm AGB}$ increasing with 
increasing metallicity (from $\eta_{\mathrm AGB}$ = 1 to $\eta_{\mathrm AGB}$ 
= 4). Furthermore, Model~3{\it n} adopts the yields from massive stars by 
Woosley \& Weaver (1995) rather than Nomoto et al. (1997). It is worth 
noticing that the yields from LIMS increase with decreasing values of 
$\eta_{\mathrm AGB}$ (i.e., smaller mass-loss rates) as a lower value of 
$\eta_{\mathrm AGB}$ results in longer AGB lifetimes and therefore more 
thermal pulses (assuming that the amount of dredged-up matter during a thermal 
pulse is roughly constant). Moreover, the yields of $^{13}$C and $^{17}$O from 
Woosley \& Weaver (1995) are larger than those from Nomoto et al. (1997), 
owing to the fact that they take into account nuclear burning outside the 
receding He-core. They also predict larger yields for $^{12}$C, but slightly 
lower ones for $^{16}$O.

As a consequence, Model~3{\it n} predicts a solar abundance of $^{12}$C larger 
than that predicted by Model~3{\it s}. On the contrary, the predicted solar 
abundance of $^{16}$O is lower (see Table~2). Therefore, the secondary 
production of $^{13}$C and $^{15}$N from novae increases, as the mean masses 
ejected from novae in the form of $^{13}$C and $^{15}$N are now scaled with a 
higher initial abundance of $^{12}$C at each time. Owing to that, higher 
$^{13}$C and $^{15}$N solar abundances are found (Table~2). On the other hand, 
the secondary production of $^{17}$O from novae in the case of Model~3{\it n} 
is lower than in the case of Model~3{\it s}, since it scales with the initial 
abundance of $^{16}$O in the star, which is now lower. However, Model~3{\it n} 
predicts for $^{17}$O a solar abundance higher than Model~3{\it s}, due to the 
non-negligible $^{17}$O contribution from massive stars when the yields of 
Woosley \& Weaver (1995) are adopted rather than those of Nomoto et al. (1997).

Once the CNO isotopic ratios are normalized to their theoretical solar values, 
it is seen that the model predictions for the $^{12}$C/$^{13}$C and 
$^{14}$N/$^{15}$N ratios from Model~3{\it n}, as far as both the temporal 
evolution at the solar ring and the present-day distribution across the disc 
are concerned, do not change appreciably with respect to those from 
Model~3{\it s} (Fig.~8 and Table~3). On the other hand, in the case of 
Model~3{\it n} the $^{16}$O/$^{17}$O ratio shows a smoother decrease from the 
time of the birth of the Sun up to now, and a shallower gradient is predicted 
across the disc. This is due to the injection into the ISM of a non-negligible 
amount of $^{17}$O from massive stars.

At this point, it has to be noticed that if the theoretical isotopic ratios 
are not normalized to the solar values predicted by the models themselves, the 
situation is even worse. Our models predict solar abundances of $^{12}$C, 
$^{14}$N and $^{16}$O in good agreement with the more recent abundance 
determinations in the solar photosphere by Holweger (2001) (except for 
$^{12}$C) and Allende Prieto, Lambert \& Asplund (2001, 2002), but too low 
when compared to the meteoritic abundances by Anders \& Grevesse (1989) 
(Table~2; see also Chiappini et al. 2003). Since the isotopic ratios given in 
Anders \& Grevesse can still be considered as the right Solar System values 
(P. Hoppe, private communication), it is likely that the solar abundances of 
the minor isotopes have to be scaled according to the revised values of the 
corresponding main isotopes. In other words, we should require an even lower 
contribution from novae and/or LIMS to the Galactic $^{13}$C and $^{17}$O 
abundances than suggested above. This may be a precious information for people 
dealing with stellar evolution and nucleosynthesis.

Finally, we address the problem of $^{17}$O destruction in stars. $^{17}$O is 
destroyed in stellar interiors by proton captures through the reactions 
$^{17}$O\,(p, $\alpha$)\,$^{14}$N and $^{17}$O\,(p, $\gamma$)\,$^{18}$F (this 
latter occurs only at high temperatures). All the models previously discussed 
assume that all the $^{17}$O injected into the ISM by stars of different 
masses is newly produced, i.e., all the $^{17}$O present in the protostellar 
nebula is destroyed in the hot stellar interior. In Fig.~9a,b we compare 
results from Model~3{\it n} {\it (solid lines)} to those from Model~3{\it m} 
{\it (dashed lines)}. Model~3{\it m} is the same as Model~3{\it n}, except for 
the fact that now all the pristine $^{17}$O is assumed to survive and to be 
returned back into the ISM at the death of the star. A realistic situation 
should be probably something in between the two. Assuming that all the 
$^{17}$O present in the gas out of which the stars form is preserved, causes a 
steepening of the $^{16}$O/$^{17}$O gradient and a more pronounced decrease of 
the ratio in the solar neighbourhood in the last 4.5 Gyr. Obviously, in the 
case of Model~3{\it m} we need also to further reduce the production of newly 
formed $^{17}$O in order not to overestimate its solar abundance. We choose to 
lower the yields of $^{17}$O from intermediate-mass stars (see Table~1).

\section{Conclusions}

In this paper we discuss the problem of the evolution of the CNO isotopes in 
both the solar vicinity and the disc of the Galaxy. In particular, we analyse 
the implications of updated results from stellar nucleosynthesis studies 
(including nova nucleosynthesis). Our main conclusions can be summarized as 
follows:
\begin{enumerate}
\item A model which considers only single low-, intermediate- and high-mass 
      stars as $^{13}$C, $^{15}$N and $^{17}$O producers, is in trouble with 
      reproducing the data relevant to the temporal variation of the CNO 
      isotopic ratios in the solar neighbourhood as well as their behavior 
      across the Galactic disc at the present time, unless one does not allow 
      for a proper revision of current stellar yields. In particular, in the 
      specific case of the carbon isotopic ratio, we show that a good 
      agreement with the observations can be obtained only by reducing the 
      strength of HBB in intermediate-mass stars.
\item Current theoretical yields of $^{17}$O from intermediate-mass stars are 
      overestimated; this may suggest the need of considering some neglected 
      reactions resulting in $^{17}$O destruction inside stars.
\item It has been claimed that novae can contribute to large amounts of 
      $^{13}$C, $^{15}$N and $^{17}$O and eventually explain the whole solar 
      abundances of these elements (e.g., Starrfield et al. 1972, 1974; 
      D'Antona \& Matteucci 1991; Matteucci \& D'Antona 1991; Woosley et al. 
      1997; Jos\'e \& Hernanz 1998). In this paper, we adopt detailed 
      nucleosynthesis in the ejecta of classical novae as published by Jos\'e 
      \& Hernanz (1998) for a grid of hydrodynamical nova models spanning a 
      wide range of CO and ONe WD masses (0.8\,--\,1.35 $M_\odot$) and mixing 
      levels between the accreted envelope and the outermost shells of the 
      underlying WD core (25\%\,--\,75\%). We find that, when included in a 
      detailed model for the chemical evolution of the Milky Way, they produce 
      $^{12}$C/$^{13}$C, $^{14}$N/$^{15}$N and $^{16}$O/$^{17}$O ratios 
      decreasing with increasing metallicity, i.e., decreasing with time at 
      the solar radius and increasing with Galactocentric distance at the 
      present time, in agreement with the trends inferred from observations. 
      However, if novae are the only $^{13}$C, $^{15}$N and $^{17}$O 
      producers, the CNO isotopic ratios are found to decrease too steeply at 
      $R = R_\odot$, often producing present-day ratios across the Galactic 
      disc lying at the lower observational limit suggested by the 
      observations at each radius. These results are almost independent of 
      whether the production of $^{13}$C, $^{15}$N and $^{17}$O from novae is 
      treated as primary or secondary. In fact, also in the case of primary 
      production, $^{13}$C, $^{15}$N and $^{17}$O behave as secondary elements 
      from a point of view of Galactic chemical evolution, owing to the long 
      delay with which they are restored into the ISM. However, in the case of 
      purely primary production (from the nucleosynthesis point of view) of 
      $^{13}$C, $^{15}$N and $^{17}$O from novae, flatter gradients and less 
      pronounced decreasing trends at the solar radius are found for all the 
      ratios.
\item A model in which $^{13}$C and $^{17}$O are produced by intermediate- and 
      high-mass stars as well as novae fits better the observations; however, 
      in the case of $^{17}$O it is necessary to lower by hand the yields from 
      both intermediate-mass stars and novae in order to reproduce the solar 
      $^{17}$O abundance. We know that $^{17}$O is always produced as a 
      secondary element in intermediate- and high-mass stars. However, our 
      models do not rule out that at least a minor fraction of the solar 
      $^{17}$O might be of primary origin. This fraction would originate from 
      nova systems.
\item The behavior of the $^{14}$N/$^{15}$N ratio along the Galactic disc and 
      as a function of time in the solar neighbourhood seems to suggest that 
      $^{15}$N has been produced on long time-scales, even longer than those 
      of $^{14}$N. Novae are the best candidates for producing $^{15}$N on 
      very long time-scales in the framework of the presently available 
      nucleosynthesis calculations. However, large uncertainties are still 
      present in the $^{15}$N abundance determinations in the local ISM and 
      this prevents us from drawing firm conclusions on the origin of this 
      element.
\end{enumerate}

\section*{Acknowledgments}

We thank C. Chiappini for many useful discussions and M. Tosi for reading the 
manuscript and giving us useful suggestions. We also acknowledge C. Allende 
Prieto, H. Holweger and P. Hoppe for explanations regarding the Solar System 
isotopic ratio data. Finally, we thank an anonymous referee for suggestions 
that improved the final version of this paper.

\bsp

\label{lastpage}


\begin{thebibliography}{99}
\bibitem{b1}
Abia C., Isern J., 1997, MNRAS, 289, L11
\bibitem{b2}
Abia C., Busso M., Gallino R., Dom\'\i nguez I., Straniero O., Isern J., 2001, 
   ApJ, 559, 1117
\bibitem{b3}
Allende Prieto C., Lambert D.L., Asplund M., 2001, ApJ, 556, L63
\bibitem{b4}
Allende Prieto C., Lambert D.L., Asplund M., 2002, ApJ, 573, L137
\bibitem{b5}
Anders E., Grevesse N., 1989, Geochim. Cosmochim. Acta, 53, 197
\bibitem{b6}
Audouze J., Lequeux J., Vigroux L., 1975, A\&A, 43, 71
\bibitem{b7}
Audouze J., Lequeux J., Rocca-Volmerange B., Vigroux L., 1977, in CNO 
   isotopes in astrophysics. D. Reidel Publishing Co., Dordrecht, p.~155
\bibitem{b8}
Bachiller R., Forveille T., Huggins P.J., Cox P., 1997, A\&A, 324, 1123
\bibitem{b9}
Balser D.S., McMullin J.P., Wilson T.L., 2002, ApJ, 572, 326
\bibitem{b10}
Bath G.T., Shaviv G., 1978, MNRAS, 183, 515
\bibitem{b11}
Boogert A.C.A. et al., 2000, A\&A, 353, 349
\bibitem{b12}
Boreiko R.T., Betz A.L., 1996, ApJ, 467, L113
\bibitem{b13}
Cameron A.G.W., 1982, in Barnes C.A., Clayton D.D., Schramm N.D., eds., Essays 
   in Nuclear Astrophysics. Cambridge Univ. Press, Cambridge, p.~23
\bibitem{b14}
Carigi L., 2000, Rev. Mex. Astron. Astrofis., 36, 171
\bibitem{b15}
Charbonnel C., do Nascimento J.D., Jr., 1998, A\&A, 336, 915
\bibitem{b16}
Charbonnel C., Brown J.A., Wallerstein G., 1998, A\&A, 332, 204
\bibitem{b17}
Chiappini C., Matteucci F., Gratton R., 1997, ApJ, 477, 765
\bibitem{b18}
Chiappini C., Matteucci F., Romano D., 2001, ApJ, 554, 1044
\bibitem{b19}
Chiappini C., Renda A., Matteucci F., 2002, A\&A, 395, 789
\bibitem{b20}
Chiappini C., Romano D., Matteucci F., 2003, MNRAS, 339, 63
\bibitem{b21}
Chin Y., Henkel C., Langer N., Mauersberger R., 1999, ApJ, 512, L143
\bibitem{b22}
Clayton D.D., Arnett D., Kane J., Meyer B.S., 1997, ApJ, 486, 824
\bibitem{b23}
Crane P., Hegyi D.J., 1988, ApJ, 326, L35
\bibitem{b24}
Dahmen G., Wilson T.L., Matteucci F., 1995, A\&A, 295, 194
\bibitem{b25}
D'Antona F., Matteucci F., 1991, A\&A, 248, 62
\bibitem{b26}
D'Antona F., Mazzitelli I., 1982, ApJ, 260, 722
\bibitem{b27}
Dearborn D., Tinsley B.M., Schramm D.N., 1978, ApJ, 223, 557
\bibitem{b28}
Della Valle M., 2000, in Matteucci F., Giovannelli F., eds., The Evolution of 
   the Milky Way: Stars versus Clusters. Kluwer Academic Publishers, p.~371
\bibitem{b29}
Della Valle M., Livio M., 1994, A\&A, 286, 786
\bibitem{b30}
Galli D., Stanghellini L., Tosi M., Palla F., 1997, ApJ, 477, 218
\bibitem{b31}
Gehrz R.D., Truran J.W., Williams R.E., Starrfield S., 1998, PASP, 110, 3
\bibitem{b32}
Gilroy K.K., Brown J.A., 1991, ApJ, 371, 578
\bibitem{b33}
Goswami A., Prantzos N., 2000, A\&A, 359, 191
\bibitem{b34}
G\"usten R., Ungerechts H., 1985, A\&A, 145, 241
\bibitem{b35}
Heger A., Woosley S.E., Langer N., 2000, NewAR, 44, 297
\bibitem{b36}
Henry R.B.C., Edmunds M.G., K\"oppen J., 2000, ApJ, 541, 660
\bibitem{b37}
Holweger H., 2001, in Wimmer-Schweingruber R.F., ed., Am. Inst. of Physics 
   Conf. Proc. Vol.~598, Joint SOHO/ACE Workshop: Solar and Galactic 
   Composition, p.~23
\bibitem{b38}
Iben I., Jr., 1964, ApJ, 140, 1631
\bibitem{b39}
Iben I., Jr., Renzini A., 1984, Phys. Letters, 105, 329
\bibitem{b40}
Jos\'e J., Hernanz M., 1998, ApJ, 494, 680
\bibitem{b41}
Keene J. et al., 1998, ApJ, 494, L107
\bibitem{b42}
Lambert D.L., Gustafsson B., Eriksson K., Hinkle K.H., 1986, ApJS, 62, 373
\bibitem{b43}
Langer W.D., Penzias A.A., 1993, ApJ, 408, 539
\bibitem{b44}
Lucas R., Liszt H., 1998, A\&A, 337, 246
\bibitem{b45}
Maeder A., 1992, A\&A, 264, 105
\bibitem{b46}
Marigo P., 2001, A\&A, 370, 194
\bibitem{b47}
Marigo P., Bressan A., Chiosi C., 1996, A\&A, 313, 545
\bibitem{b48}
Matteucci F., 1986, MNRAS, 221, 911
\bibitem{b49}
Matteucci F., D'Antona F., 1991, A\&A, 247, L37
\bibitem{b50}
Matteucci F., Fran\c cois P., 1989, MNRAS, 239, 885
\bibitem{b51}
Meynet G., Maeder A., 2002a, A\&A, 390, 561
\bibitem{b52}
Meynet G., Maeder A., 2002b, A\&A, 381, L25
\bibitem{b53}
Nomoto K., Hashimoto M., Tsujimoto T., Thielemann F.-K., Kishimoto N., 
   Kubo Y., Nakasato N., 1997, Nucl. Phys. A, 616, 79c
\bibitem{b54}
Nomoto K., Thielemann F.-K., Yokoi K., 1984, ApJ, 286, 644
\bibitem{b55}
Ohnaka K., Tsuji T., 1996, A\&A, 310, 933
\bibitem{b56}
Ohnaka K., Tsuji T., 1999, A\&A, 345, 233
\bibitem{b57}
Ohnaka K., Tsuji T., Aoki W., 2000, A\&A, 353, 528
\bibitem{b58}
Palla F., Bachiller R., Stanghellini L., Tosi M., Galli D., 2000, A\&A, 355, 69
\bibitem{b59}
Politano M., Starrfield S., Truran J.W., Weiss A., Sparks W.M., 1995, ApJ, 
   448, 807
\bibitem{b60}
Prantzos N., Vangioni-Flam E., Chauveau S., 1994, A\&A, 285, 132
\bibitem{b61}
Prantzos N., Aubert O., Audouze J., 1996, A\&A, 309, 760
\bibitem{b62}
Renzini A., Voli M., 1981, A\&A, 94, 175
\bibitem{b63}
Romano D., Matteucci F., Molaro P., Bonifacio P., 1999, A\&A, 352, 117
\bibitem{b64}
Romano D., Matteucci F., Ventura P., D'Antona F., 2001, A\&A, 374, 646
\bibitem{b65}
Savage C., Apponi A.J., Ziurys L.M., 2001, AAS, 198, 5913
\bibitem{b66}
Scalo J.M., 1986, Fund. Cosmic Phys., 11, 1
\bibitem{b67}
Sch\"oier F.L., Olofsson H., 2000, A\&A, 359, 586
\bibitem{b68}
Shafter A.W., 1997, ApJ, 487, 226
\bibitem{b69}
Shara M.M., Zurek D.R., Williams R.E., Prialnik D., Gilmozzi R., Moffat 
   A.F.J., 1997, AJ, 114, 258
\bibitem{b70}
Sheffer Y., Lambert D.L., Federman S.R., 2002, ApJ, 574, L171
\bibitem{b71}
Smith V.V., Terndrup D.M., Suntzeff N.B., 2002, ApJ, 579, 832
\bibitem{b72}
Stahl O., Appenzeller I., Wilson T.L., Henkel C., 1989, A\&A, 221, 321
\bibitem{b73}
Starrfield S., Truran J.W., Sparks W.M., Kutter G.S., 1972, ApJ, 176, 169
\bibitem{b74}
Starrfield S., Sparks W.M., Truran J.W., 1974, ApJ, 192, 647
\bibitem{b75}
Talbot R.J., Arnett D.W., 1974, ApJ, 190, 605
\bibitem{b76}
Thielemann F.-K., Nomoto K., Hashimoto M., 1993, in Prantzos N., Vangioni-Flam 
   E., Cass\'e M., eds., Origin and Evolution of the Elements. Cambridge Univ. 
   Press, Cambridge, p.~297 
\bibitem{b77}
Timmes F.X., Woosley S.E., Weaver T.A., 1995, ApJS, 98, 617
\bibitem{b78}
Tosi M., 1982, ApJ, 254, 699
\bibitem{b79}
Tosi M., 1988, A\&A, 197, 33
\bibitem{b80}
Tosi M., 2000, in Matteucci F., Giovannelli F., eds., The Evolution of the 
   Milky Way: Stars versus Clusters. Kluwer Academic Publishers, p.~505
\bibitem{b81}
van den Hoek L.B., Groenewegen M.A.T., 1997, A\&AS, 123, 305
\bibitem{b82}
Ventura P., D'Antona F., Mazzitelli I., 2002, A\&A, 393, 215
\bibitem{b83}
Vigroux L., Audouze J., Lequeux J., 1976, A\&A, 52, 1
\bibitem{b84}
Wannier P.G., Linke R.A., Penzias A.A., 1981, ApJ, 247, 522
\bibitem{b85}
Wilson T.L., Rood R.T., 1994, ARA\&A, 32, 191
\bibitem{b86}
Woosley S.E., Weaver T.A., 1995, ApJS, 101, 181
\bibitem{b87}
Woosley S.E., Hoffman R.D., Timmes F.X., Weaver T.A., Thielemann F.-K., 1997, 
   Nucl. Phys. A, 621, 445c
\end{thebibliography}
\end{document}